\documentclass[reqno]{amsart}
\usepackage{hyperref}
\usepackage{amsmath,amssymb,graphicx}

\AtBeginDocument{{\noindent\small
\emph{Journal of Fractional Calculus and Applications}\\
Vol. 10(2) July    2019,  pp. 92-135\newline
ISSN: 2090-5858.\\
http://math-frac.oreg/Journals/JFCA/\\
------------------------------------------------------------------------------------------------}
\vspace{9mm}}

\begin{document}
\title[\hfilneg JFCA-2019/10(2)\hfil On the nature of the conformable derivative]
{ On the nature of the conformable derivative and
its applications to physics}

\author[D. R. Anderson, E. Camrud, D. J. Ulness \hfil JFCA-2019/10(2)\hfilneg]
{D. R. Anderson, E. Camrud, D. J. Ulness}

\address{Douglas R. Anderson \newline
Department of Mathematics, Concordia College, Moorhead MN 56562, USA}
\email{andersod@cord.edu}

\address{Evan Camrud  \newline
Department of Mathematics, Iowa State University, Ames, IA 50011, USA}
\email{ecamrud@iastate.edu}

\address{Darin J. Ulness \newline
Department of Chemistry, Concordia College, Moorhead MN 56562, USA}
\email{ulnessd@cord.edu}

\thanks{Submitted   Aug.   22, 2018. }

\setcounter{page}{92}

\begin{abstract}
The purpose of this work is to show that the Khalil and Katagampoula
conformable derivatives are equivalent to the simple change of variables $x$ 
$\rightarrow $ $x^{\alpha }/\alpha ,$ where $\alpha $ is the order of the
derivative operator, when applied to differential functions. Although this
means no \textquotedblleft new mathematics\textquotedblright\ is obtained by
working with these derivatives, it is a second purpose of this work to argue
that there is still significant value in exploring the mathematics and
physical applications of these derivatives. This work considers linear
differential equations, self-adjointness, Sturm-Liouville systems, and
integral transforms. A third purpose of this work is to contribute to the
physical interpretation when these derivatives are applied to physics and
engineering. Quantum mechanics serves as the primary backdrop for this
development.
\end{abstract}

\maketitle

\section{Introduction}

The concept of a fractional derivative has been receiving a lot of attention
in the literature in recent years,\cite%
{West,Machado,Machado2,Oldham,Herrmann} with entire journals devoted to
fractional analysis \cite{Journals}. Many of the authors of these papers
mention the famous correspondence between Leibniz and L'H\^{o}pital in 1695.
Over the intervening years many definitions of a fractional derivative have
appeared; well known examples being the Riemann-Liouville and the Caputo
definitions. The current activity clearly suggests the extension of
derivatives of non-integer power is not straightforward to say the least.

In fact, the defining properties of such derivatives are not agreed upon.
It is typical that a particular definition captures only some of the
properties of the conventional derivative. Ortigueira and Machado have
recently compared and contrasted definitions of fractional derivatives and
have set forth criteria for such derivatives \cite{Ortigueira}. This has led
to some definitions of fractional derivatives to be reclassified as
conformable derivatives. Zhao and Luo \cite{ZhaoLuo} provide a good account
heredity/nonhereditary and locality/nonlocality \cite{Uchaikin}. In a very
recent work, Tarasov clearly discusses nonlocality in the context of a
number of familiar fractional derivatives including the Khalil and
Katugampola definitions, which are the focus of this current paper.\cite%
{Tarasov} Tarasov points out that equations involving these two conformable
derivatives can be reduced to ordinary differential equations. The current
paper elaborates on that assertion.

In 2014 Khalil suggested the definition \cite{Khalil},

\begin{eqnarray}
D^{\alpha }[f(x)] &=&\lim_{\epsilon \rightarrow 0}\frac{f(x+\epsilon
x^{1-\alpha })-f(x)}{\epsilon },\quad x>0 \\
D^{\alpha }[f(0)] &=&\lim_{t\rightarrow 0^{+}}D^{\alpha }[f(x)].  \notag
\end{eqnarray}%
Katugampola shortly thereafter worked out a few additional technical details %
\cite{Udita01,Udita02}. For brevity we shall refer to the above derivative as simply
the conformable derivative in this work. The conformable derivative was subsequently
generalized in several ways \cite{Abdel}. Many papers have appeared based on
exploring properties \cite{Abdel,Atangana,Doug2,Bayour,Barham} and physical
applications \cite{Doug,Yucel,Hosseini,Neto,Karayer,ZhaoYang,YangWang} of the
conformable derivative.

A case of particular interest, particularly with an eye toward applications
in physics and engineering, is applying the conformable derivative to differentiable
functions. In this case the conformable derivative becomes

\begin{equation}
D^{\alpha }[f(x)]=x^{1-\alpha }\frac{df(x)}{dx}.  \label{KKder}
\end{equation}

In operator language,

\begin{equation}
D^{\alpha }\equiv x^{1-\alpha }\frac{d}{dx}.  \label{KKderop}
\end{equation}

This leads to the main point of the current work: \emph{The conformable derivative
for differentiable functions is equivalent to a simple change of variable}.
Precisely, $u=x^{\alpha }/\alpha $. It should be noted that a criticism of
the conformable derivative is that, although conformable at the limit $\alpha
\rightarrow 1$ ($\lim_{\alpha \rightarrow 1}D^{\alpha }f=f^{\prime }$), it
is not conformable at the other limit, $\alpha \rightarrow 0,$ ($%
\lim_{\alpha \rightarrow 0}D^{\alpha }f\neq f$). From the point of view of
the assertion about the equality of the conformable derivative to a change of
variables, one can say that the conformable derivative is not conformable as $\alpha
\rightarrow 0$ because $t^{\alpha }/\alpha $ is undefined at $\alpha =0.$

As such, the conformable derivative does not contribute \textquotedblleft new
mathematics.\textquotedblright\ That said, exploration of the conformable derivative
and its generalizations can still be interesting and valuable. The focus of
this paper is to elucidate the nature of this change of variables in a
variety of settings. Further, we hope to provide some physical insight to
assist with use in the applied setting. We focus on application in quantum
mechanics but some of the discussion about how to interpret physical units
and spaces related via Fourier transformation are relevant in general.

First, the basic calculus of the conformable derivative is laid out. Second,
self-adjointness is discussed and the Sturm-Liouville system under this
change of variable is presented with examples. Finally, integral transforms,
specifically the Fourier and Laplace transforms, are discussed. The
interpretation of the meaning of the physical units is presented in context
of each of these settings. Application to quantum mechanics follows the
mathematical development and proceeds concluding remarks. For ease of
discussion, the word \textquotedblleft conformable\textquotedblright\ will
be used as an adjective to describe the conformable derivative type of change of
variable. For example, \textquotedblleft conformable Bessel
function,\textquotedblright\ \textquotedblleft conformable Laplace
transform\textquotedblright\ etc. This does not imply that there is
something fundamentally new about a conformable object compared to its
standard counter-part. \emph{They are always related via a simple change of
variable.}

\section{The basic calculus of the conformable derivative}

In this section the equivalence of calculus of differentiable functions
using the conformable derivative and the change of variable $u=x^{\alpha }/\alpha $
is demonstrated.

The conformable derivative has the following important properties. This definition
yields the following results (from Theorem 2.3 of Katugampola \cite{Udita01})

\begin{itemize}
\item $D^{\alpha }[af+bg]=aD^{\alpha }[f]+bD^{\alpha }[g]$ (linearity).

\item $D^{\alpha }[fg]=fD^{\alpha }[g]+gD^{\alpha }[f]$ (product rule).

\item $D^{\alpha }[f(g)]=\frac{df}{dg}D^{\alpha }[g]$ (chain rule).

\item $D^{\alpha }[f]=x^{1-\alpha }f^{\prime },\quad \mbox{where}\quad
f^{\prime }=\frac{df}{dx}$.
\end{itemize}

To see the equivalence of the conformable derivative and the change of variables $%
u=x^{\alpha }/\alpha ,$ consider direct substitution and the chain rule
in 
\begin{equation}
D^{\alpha }f(x)\equiv x^{1-\alpha }\frac{df(x)}{dx}.
\end{equation}%
Then, 
\begin{equation}
x^{1-\alpha }\frac{df(x)}{dx}=x^{1-\alpha }\frac{df(u)}{du}\frac{du}{dx}%
=x^{1-\alpha }\frac{df(u)}{du}x^{\alpha -1}=\frac{df(u)}{du}.
\end{equation}

\subsection{Second order linear differential equation}

Consider the general second order linear differential equation (SOLDE)%
\begin{equation}
p(u)\frac{d^{2}y(u)}{du^{2}}+q(u)\frac{dy(u)}{du}+r(u)y(u)=s(u).
\label{SOLDE}
\end{equation}%
Now let 
\begin{equation}
u=\frac{x^{\alpha }}{\alpha },\quad du=x^{\alpha -1}dx.
\end{equation}%
Thus, 
\begin{equation}
\frac{dy(u)}{du}=\frac{dy\left( \frac{x^{\alpha }}{\alpha }\right) }{dx}%
\frac{dx}{du}=x^{1-\alpha }\frac{dy\left( \frac{x^{\alpha }}{\alpha }\right) 
}{dx}=D^{\alpha }y\left( \frac{x^{\alpha }}{\alpha }\right) 
\end{equation}%
and%
\begin{equation}
\frac{d^{2}y(u)}{du^{2}}=x^{2-2\alpha }\frac{d^{2}y}{dx^{2}}+\left( 1-\alpha
\right) x^{1-2\alpha }\frac{dy}{dx}\equiv \hat{C}_{2\alpha }y\left( \frac{%
x^{\alpha }}{\alpha }\right) .
\end{equation}%
Therefore Eq. (\ref{SOLDE}) becomes%
\begin{eqnarray}
p\left( \frac{x^{\alpha }}{\alpha }\right) \hat{C}_{2\alpha }y\left( \frac{%
x^{\alpha }}{\alpha }\right) +q\left( \frac{x^{\alpha }}{\alpha }\right)
D_{x}^{\alpha }y\left( \frac{x^{\alpha }}{\alpha }\right) &&  \label{fSOLDE1}
\\
+r\left( \frac{x^{\alpha }}{\alpha }\right) y\left( \frac{x^{\alpha }}{%
\alpha }\right) &=&s\left( \frac{x^{\alpha }}{\alpha }\right) .  \notag
\end{eqnarray}%
This provides a recipe for translating any normal SOLDE\ into a conformable
SOLDE. The notion of a \textquotedblleft natural\textquotedblright\
variable, $\frac{x^{\alpha }}{\alpha },$ for the conformable derivative
arises. The simple change of variable $x\longleftrightarrow \frac{x^{\alpha }%
}{\alpha }$ pulls all SOLDEs into conformable SOLDEs and vice versa.

It is interesting to expand Eq. (\ref{fSOLDE1}). 
\begin{eqnarray}
p\left( \frac{x^{\alpha }}{\alpha }\right) \hat{C}_{2\alpha }y+q\left( \frac{%
x^{\alpha }}{\alpha }\right) D_{x}^{\alpha }y+r\left( \frac{x^{\alpha }}{%
\alpha }\right) y &=&s\left( \frac{x^{\alpha }}{\alpha }\right)  \notag \\
p\left( \frac{x^{\alpha }}{\alpha }\right) \left( x^{2-2\alpha }y^{\prime
\prime }+\left( 1-\alpha \right) x^{1-2\alpha }y^{\prime }\right) +q\left( 
\frac{x^{\alpha }}{\alpha }\right) x^{1-\alpha }y^{\prime }  \notag\\
+r\left( \frac{%
x^{\alpha }}{\alpha }\right) y &=&s\left( \frac{x^{\alpha }}{\alpha }\right)
\notag \\
x^{2-2\alpha }p\left( \frac{x^{\alpha }}{\alpha }\right) y^{\prime \prime
}+\left( \left( 1-\alpha \right) x^{1-2\alpha }p\left( \frac{x^{\alpha }}{%
\alpha }\right) +x^{1-\alpha }q\left( \frac{x^{\alpha }}{\alpha }\right)
\right) y^{\prime }  \notag\\
+r\left( \frac{x^{\alpha }}{\alpha }\right) y &=&s\left( 
\frac{x^{\alpha }}{\alpha }\right)  \notag \\
P(x)y^{\prime \prime }+Q(x)y^{\prime }+R(x)y &=&S(x),  \label{fSOLDE2}
\end{eqnarray}%
where 
\begin{eqnarray}
P(x) &=&p\left( \frac{x^{\alpha }}{\alpha }\right) x^{2-2\alpha }  \notag \\
Q(x) &=&\left( 1-\alpha \right) x^{1-2\alpha }p\left( \frac{x^{\alpha }}{%
\alpha }\right) +x^{1-\alpha }q\left( \frac{x^{\alpha }}{\alpha }\right) 
\notag \\
R(x) &=&r\left( \frac{x^{\alpha }}{\alpha }\right)  \notag \\
S(x) &=&s\left( \frac{x^{\alpha }}{\alpha }\right) .  \label{convdefs}
\end{eqnarray}

\noindent\textbf{Example 1}

Let's consider some examples. First Bessel's equation%
\begin{equation}
u^{2}\frac{d^{2}y(u)}{du^{2}}+u\frac{dy(u)}{du}+\left( u^{2}+v^{2}\right)
y(u)=0,
\end{equation}%
which has solution 
\begin{equation}
y(u)=C_{1}J_{v}(u)+C_{2}Y_{v}(u).
\end{equation}%
The corresponding conformable Bessel's function according to the recipe is 
\begin{eqnarray}
\left( \frac{x^{\alpha }}{\alpha }\right) ^{2}\hat{C}_{2\alpha }y\left( 
\frac{x^{\alpha }}{\alpha }\right) +\left( \frac{x^{\alpha }}{\alpha }%
\right) D_{x}^{\alpha }y\left( \frac{x^{\alpha }}{\alpha }\right) && \\
+\left( \left( \frac{x^{\alpha }}{\alpha }\right) ^{2}-v^{2}\right) y\left( 
\frac{x^{\alpha }}{\alpha }\right) &=&0,
\end{eqnarray}%
which has solution 
\begin{equation}
y\left( \frac{x^{\alpha }}{\alpha }\right) =C_{1}J_{v}\left( \frac{x^{\alpha
}}{\alpha }\right) +C_{2}Y_{v}\left( \frac{x^{\alpha }}{\alpha }\right) .
\end{equation}%
The expanded SOLDE becomes via 
\begin{eqnarray}
P(x) &=&\left( \frac{x^{\alpha }}{\alpha }\right) ^{2}x^{2-2\alpha }=\frac{%
x^{2}}{\alpha ^{2}}  \notag \\
Q(x) &=&\left( 1-\alpha \right) x^{1-2\alpha }\left( \frac{x^{\alpha }}{%
\alpha }\right) ^{2}+x^{1-\alpha }\left( \frac{x^{\alpha }}{\alpha }\right) =%
\frac{x}{\alpha ^{2}}  \notag \\
R(x) &=&\left( \left( \frac{x^{\alpha }}{\alpha }\right) ^{2}-v^{2}\right) 
\notag \\
S(x) &=&0.
\end{eqnarray}%
So one obtains 
\begin{eqnarray}
\frac{x^{2}}{\alpha ^{2}}y^{\prime \prime }+\frac{x}{\alpha ^{2}}y^{\prime
}+\left( \left( \frac{x^{\alpha }}{\alpha }\right) ^{2}-v^{2}\right) y &=&0 
\notag \\
x^{2}y^{\prime \prime }+xy^{\prime }+\left( x^{2\alpha }-\alpha
^{2}v^{2}\right) y &=&0.
\end{eqnarray}

\noindent\textbf{Example 2}

Consider the differential equation for confluent hypergeometric limit
function%
\begin{equation}
u\frac{d^{2}y(u)}{du^{2}}+b\frac{dy(u)}{du}-y(u)=0.
\end{equation}%
This has solution%
\begin{equation}
y(u)=C_{1}\,_{0}F_{1}(;a;u)+C_{2}u^{1-b}\,_{0}F_{1}(;2-a;u).
\end{equation}%
So%
\begin{equation}
\frac{x^{\alpha }}{\alpha }\hat{C}_{2\alpha }y+bD_{x}^{\alpha }y-y=0
\end{equation}%
has solutions 
\begin{equation}
y(x)=C_{1}\,_{0}F_{1}(;b;\frac{x^{\alpha }}{\alpha })+C_{2}\left( \frac{%
x^{\alpha }}{\alpha }\right) ^{1-b}\,_{0}F_{1}(;2-b;\frac{x^{\alpha }}{%
\alpha }).
\end{equation}%
Here 
\begin{eqnarray}
P(x) &=&\left( \frac{x^{\alpha }}{\alpha }\right) x^{2-2\alpha }=\frac{%
x^{2-\alpha }}{\alpha }  \notag \\
Q(x) &=&\left( 1-\alpha \right) x^{1-2\alpha }\left( \frac{x^{\alpha }}{%
\alpha }\right) +bx^{1-\alpha }=\left( \frac{1}{\alpha }-1+b\right)
x^{1-\alpha }  \notag \\
R(x) &=&-1  \notag \\
S(x) &=&0,
\end{eqnarray}%
so,%
\begin{equation}
\frac{x^{2-\alpha }}{\alpha }y^{\prime \prime }+\left( \frac{1}{\alpha }%
-1+b\right) x^{1-\alpha }y^{\prime }-y=0.
\end{equation}

\noindent\textbf{Example 3}

Finally, consider Airy's differential equation%
\begin{equation}
y^{\prime \prime }-xy=0,
\end{equation}%
which has solutions 
\begin{equation}
y=C_{1}\text{Ai}[x]+C_{2}\text{Bi}[x].
\end{equation}%
So,%
\begin{equation}
\hat{C}_{2\alpha }y-\frac{x^{\alpha }}{\alpha }y=0
\end{equation}%
has solutions 
\begin{equation}
y=C_{1}\text{Ai}\left[ \frac{x^{\alpha }}{\alpha }\right] +C_{2}\text{Bi}%
\left[ \frac{x^{\alpha }}{\alpha }\right].
\end{equation}

\section{Self adjointness and Sturm-Liouville systems}

Several properties of $D^{\alpha }$ have recently been investigated.\cite%
{Doug} In that work the conformable analogue of $D^{2}=\frac{d^{2}}{dx^{2}}$
was developed by first simply considering $D^{\alpha }D^{\alpha }$ (which is
not equal to $D^{2\alpha }$). This, however, is not self-adjoint but can be
made so by standard methods \cite{Hassani}. Doing so results in the
self-adjoint operator \cite{Doug} 
\begin{equation}
\hat{A}_{2\alpha }=\frac{d}{dx}\left[ x^{1-\alpha }\frac{d}{dx}\right] ,
\label{selfadjAop}
\end{equation}%
note that $\lim_{\alpha \rightarrow 1}\hat{A}{_{2\alpha }=D^{2}.}$ The
eigenvalue equation 
\begin{equation}
\hat{A}_{2\alpha }y+E_{n}y=0  \label{eigenA}
\end{equation}%
was solved in reference \cite{Doug}. This is the simplest conformable
Sturm-Liouville system and its solutions are thoroughly explored in the
remainder of this section. Several other Sturm-Liouville systems are
discussed more briefly at the end of this section. For the special case of
boundary conditions $y(0)=y(1)=0$, the (normalized) solutions are 
\begin{equation}
y=\mathbb{J}_{n}^{(\alpha )}(x)\equiv \frac{\sqrt{x^{\alpha }}J_{\eta
}\left( n_{\eta }(x^{\alpha })^{\frac{1}{2\eta }}\right) }{\sqrt{(\eta
-1)J_{\eta -1}(n_{\eta })J_{\eta +1}(n_{\eta })}},  \label{bigJ}
\end{equation}%
where $J$ is Bessel's function, $\eta =\frac{\alpha }{1+\alpha }$ and $%
n_{\eta }$ is the $n^{th}$ zero of $J_{\eta }(x)$. The eigenvalues are 
\begin{equation}
E_{n}=\frac{(1+\alpha )^{2}n_{\eta }^{2}}{4}.  \label{eigenvals}
\end{equation}%
The first three ($n=1,$ 2, 3) $\mathbb{J}_{n}^{(\alpha )}$ are plotted in
Fig 1 for $\alpha =1/4,$ 1/2, 3/4, and 1.

\begin{figure}
\includegraphics[trim=75  275 75 175,clip, width=\textwidth]{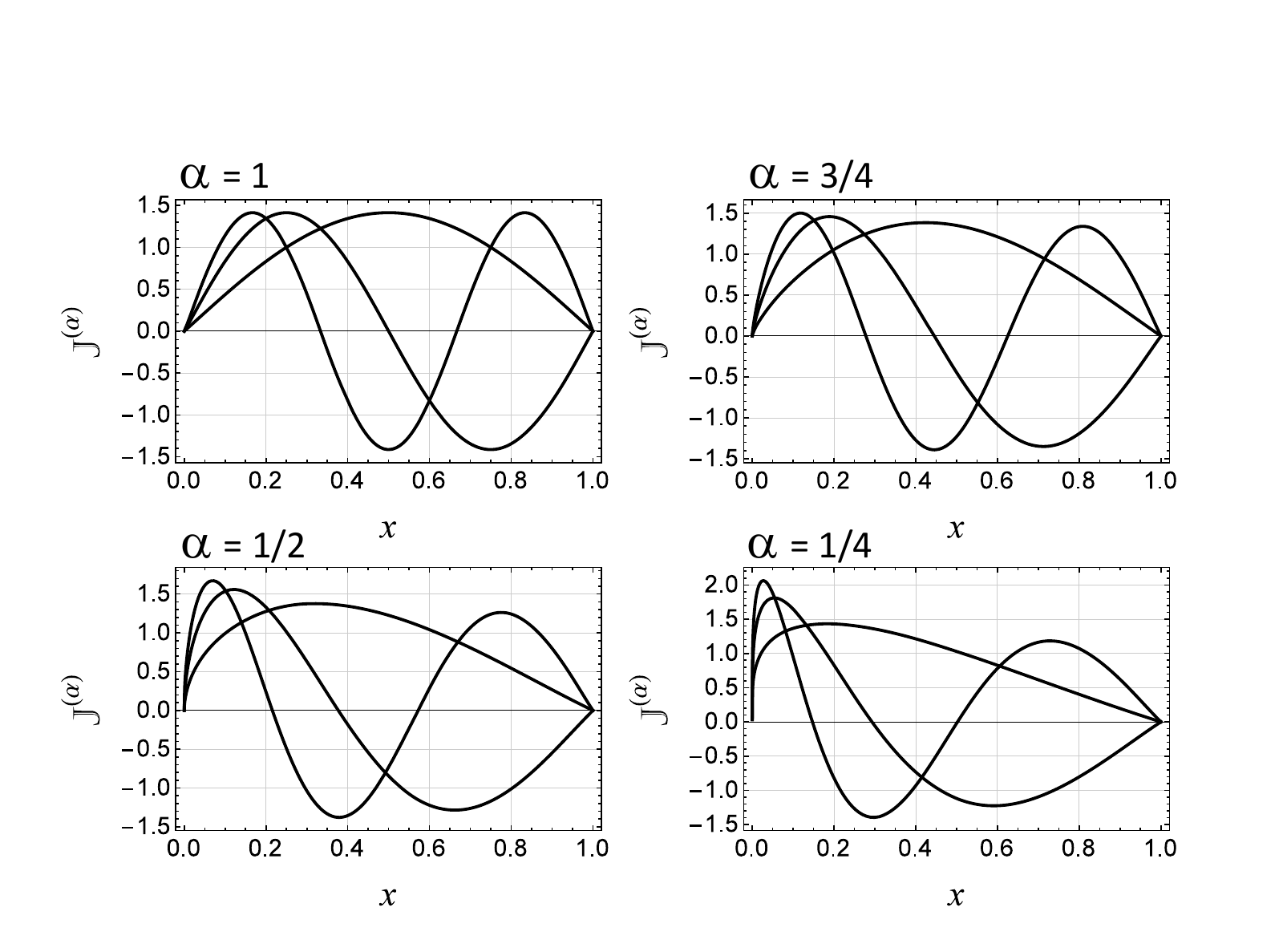}
\caption{ Each panel shows plots of $\mathbb{J}_{1}^{(\alpha )},$ $%
\mathbb{J}_{2}^{(\alpha )},$ and $\mathbb{J}_{3}^{(\alpha )}$ for $\alpha =1,
$ $3/4,$ $1/2,$ and $1/4$ (clockwise from the top left panel). These
functions are solutions of Eq. (\ref{eigenA}) with boundary conditions $%
\mathbb{J}_{n}^{(\alpha )}(0)=\mathbb{J}_{n}^{(\alpha )}(1)=0\ $and they
form a complete orthonormal set. The curves in each panel can be identified
by the fact that $\mathbb{J}_{n}^{(\alpha )}$ has $n-1$ zeros between $0<x<1.
$ The most noticeable characteristic of the $\mathbb{J}_{n}^{(\alpha )}$
functions is the skewing towards lower values of $x$ as $\alpha $ decreases.}
\end{figure}

The $\mathbb{J}_{n}^{(\alpha )}$ functions form a complete, orthonormal set
over the domain $0\leq x\leq 1$ and are a generalization of the set of
orthonormal sine functions over the same domain, $y_{n}=\sqrt{2}\sin {n\pi x}
$. The $\mathbb{J}_{n}^{(\alpha )}$ functions serve to introduce a
parameterized (by $\alpha $) extension of the harmonic functions in a manner
that has a bit more richness than a Fourier-Bessel series.

Several aspects of the $\mathbb{J}_{n}^{(\alpha )}$ functions are now
investigated. This is offered as an example of how results arising from conformable
derivative based equations can still offer interesting subject matter to
study despite the fact that the results can be obtained via a simple change
of variable. Many relations can be obtained analytically but often one must
resort to numerical calculations. The properties of the zeros of $\mathbb{J}%
_{n}^{(\alpha )}$ and a scaling factor for the $\mathbb{J}_{n}^{(\alpha )}$
functions are first explored. Then the expansion of an arbitrary function is
investigated with some representative examples.

\begin{figure}
\includegraphics[trim=75  90 75 50,clip, width=\textwidth]{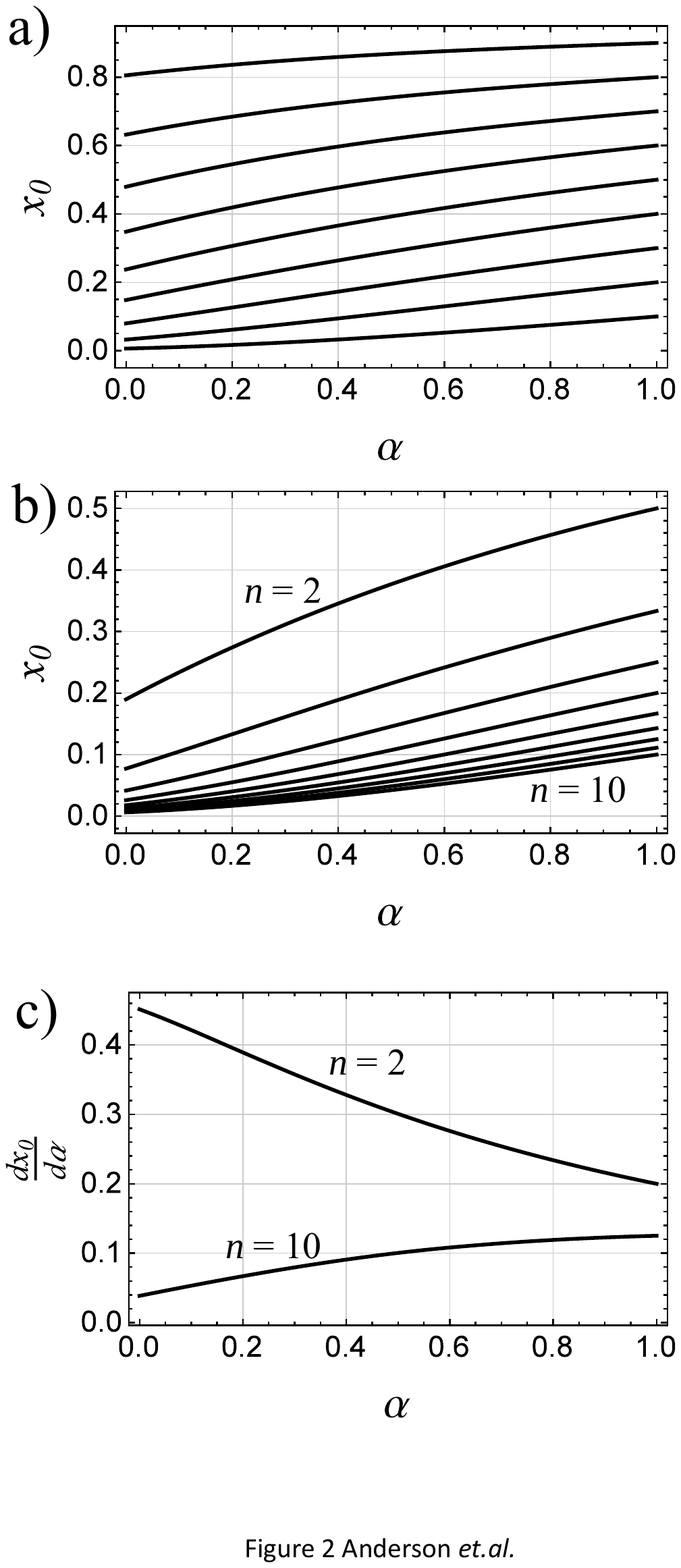}
\caption{The position of the $x=x_{0}$ zeros of $\mathbb{J}%
_{n}^{(\alpha )}.$ \textbf{(a)} The position of the first 9 zeros of $%
\mathbb{J}_{10}^{(\alpha )}.$ Each of these zeros decreases with decreasing $%
\alpha $. \textbf{(b)} The position of the first zeros of $\mathbb{J}%
_{2}^{(\alpha )}$ through $\mathbb{J}_{10}^{(\alpha )}.$ The position of
these zeros also decrease with decreasing $\alpha $ but not in the same
fashion. \textbf{(c)} The derivative of the top ($n=2$) and bottom ($n=10$)
curves in (b).}
\end{figure}

\subsection{Zeros and scaling of $\mathbb{J}_{n}^{(\protect\alpha )}$}

The position of the zeros of $\mathbb{J}_{n}^{(\alpha )}$ are determined by
a combination of the particular Bessel function involved and by the $%
x^{\alpha }$ appearing in its argument. The position of the $k^{th}$ zero of 
$\mathbb{J}_{n}^{(\alpha )},$ $\mathbb{N}_{n}^{(\alpha )}(k),$ is given by
the formula%
\begin{equation}
\mathbb{N}_{n}^{(\alpha )}(k)=\left( \frac{k_{\eta }}{n_{\eta }}\right) ^{%
\frac{2}{1+\alpha }}.  \label{bigJzeros}
\end{equation}%
Figure 2 shows $\mathbb{N}_{n}^{(\alpha )}(k)$ for a variety of different $n$
and $k$ values as a function of $\alpha .$ Figure 2a shows the positions of
the nine zeros of $\mathbb{J}_{10}^{(\alpha )}$ as a function of $\alpha $
and Fig. 2b shows the position of the first zero for $\mathbb{J}%
_{n}^{(\alpha )}$ where $n=2,\ldots ,10$. The limits of $\mathbb{N}%
_{n}^{(\alpha )}(k)$ are 
\begin{equation}
\lim_{\alpha \rightarrow 1}\mathbb{N}_{n}^{(\alpha )}(k)=\frac{k_{\frac{1}{2}%
}}{n_{\frac{1}{2}}}=\frac{k}{n}
\end{equation}%
and%
\begin{equation}
\lim_{\alpha \rightarrow 0}\mathbb{N}_{n}^{(\alpha )}(k)=\frac{k_{0}^{2}}{%
n_{0}^{2}}.
\end{equation}%
Although the graphs shown in Fig.2 are simply manifestations of the
properties of the zeros of the Bessel function, it is insightful to point
out some features. All zeros move to smaller values as $\alpha $ decreases
but do so along different trajectories such that the spacing between zeros
is the same for $\alpha =1$ but is increasing for $\alpha <1.$ Figure 2c
shows the derivative of the $\mathbb{J}_{2}^{(\alpha )}$ and $\mathbb{J}%
_{10}^{(\alpha )}$ curves of Fig. 2b. These exhibit opposite behavior with $%
\alpha .$

In much the same way that $y_{1}=\sin \pi x$ can be scaled to $y_{2}=\sin
2\pi x$ by letting $x\rightarrow 2x,$ one can determine a scaling factor, $%
s, $ such that $\mathbb{J}_{1}^{(\alpha )}(sx)\propto \mathbb{J}%
_{2}^{(\alpha )}(x).$ A second amplitude scaling factor, $N_s$, is needed to
create the equality, $N_s \mathbb{J}_{1}^{(\alpha )}(sx)=\mathbb{J}%
_{2}^{(\alpha )}(x).$ More generally, the scaling factors such that $N_s \mathbb{J}_{n}^{(\alpha )}(sx)=\mathbb{J}_{n+1}^{(\alpha )}(x)$ are,%
\begin{equation}
s=1/\mathbb{N}_{n+1}^{(\alpha )}(n)
\end{equation}%
and%
\begin{equation}
N_s =\sqrt{\frac{1}{\sqrt{s}}\cdot \frac{J_{\eta -1}(n_{\eta })J_{\eta
+1}(n_{\eta })}{J_{\eta -1}((n+1)_{\eta })J_{\eta +1}((n+1)_{\eta })}}.
\end{equation}

\subsection{Integrals of $\mathbb{J}_{n}^{(\protect\alpha )}$}

\begin{figure}
\includegraphics[trim=75  175 75 150,clip, width=\textwidth]{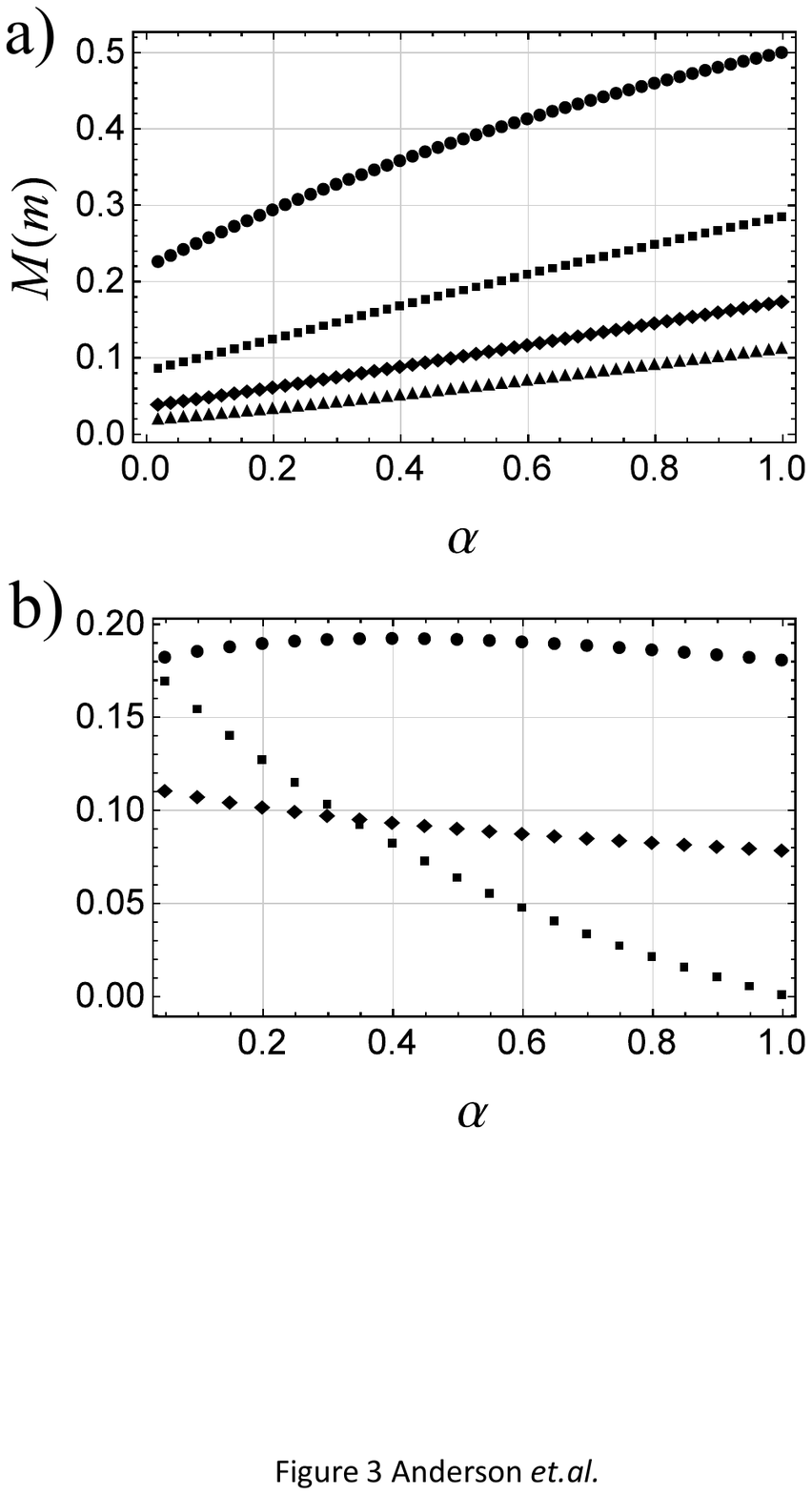}
\caption{ The \textbf{(a)} first four moments ($\bullet $ 1, $\blacksquare $ 2,
 $\blacklozenge $ 3, $\blacktriangle $ 4) and\textbf{\ (b)}
the standard deviation ($\bullet $), skewness ($\blacksquare $) and kurtosis
($\blacklozenge $) for $\left( \mathbb{J}_{1}^{(\alpha )}\right) ^{2}.$ The
standardized moment (rather than the cumulant) definition is being used for
skewness and kurtosis. The mean and skewness are relatively strong functions
of $\alpha $ whereas the standard deviation and kurtosis are weak functions
of $\alpha .$}
\end{figure}

Most of the integrals involving $\mathbb{J}_{n}^{(\alpha )}$ need to be
evaluated numerically, including showing orthonormality in the general case.
One important class of integrals is the moments of $\left( \mathbb{J}%
_{n}^{(\alpha )}\right) ^{2}$,%
\begin{equation}
M(m)=\int_{0}^{1}x^{m}\left( \mathbb{J}_{n}^{(\alpha )}\right) ^{2}.
\end{equation}%
Figure 3a shows the first four moments of $\left( \mathbb{J}_{1}^{(\alpha
)}\right) ^{2}$ as a function of $\alpha $ and Fig. 3b shows the standard
deviation, skewness, and kurtosis of $\left( \mathbb{J}_{1}^{(\alpha
)}\right) ^{2}$ also as a function of $\alpha .$ The standard deviation and
the kurtosis are weak functions of $\alpha $. The kurtosis increases
slightly with decreasing $\alpha $ but the standard deviation slightly
increases with decreasing $\alpha $ until about $\alpha =2/5$ then it
slightly decreases as $\alpha $ tends to zero. The skewness on the other
hand is a stronger function of $\alpha $ as it rises from zero at $\alpha =1$
to approximately 0.18 as $\alpha \rightarrow 0.$

\begin{figure}
\includegraphics[trim=75  250 75 75,clip, width=\textwidth]{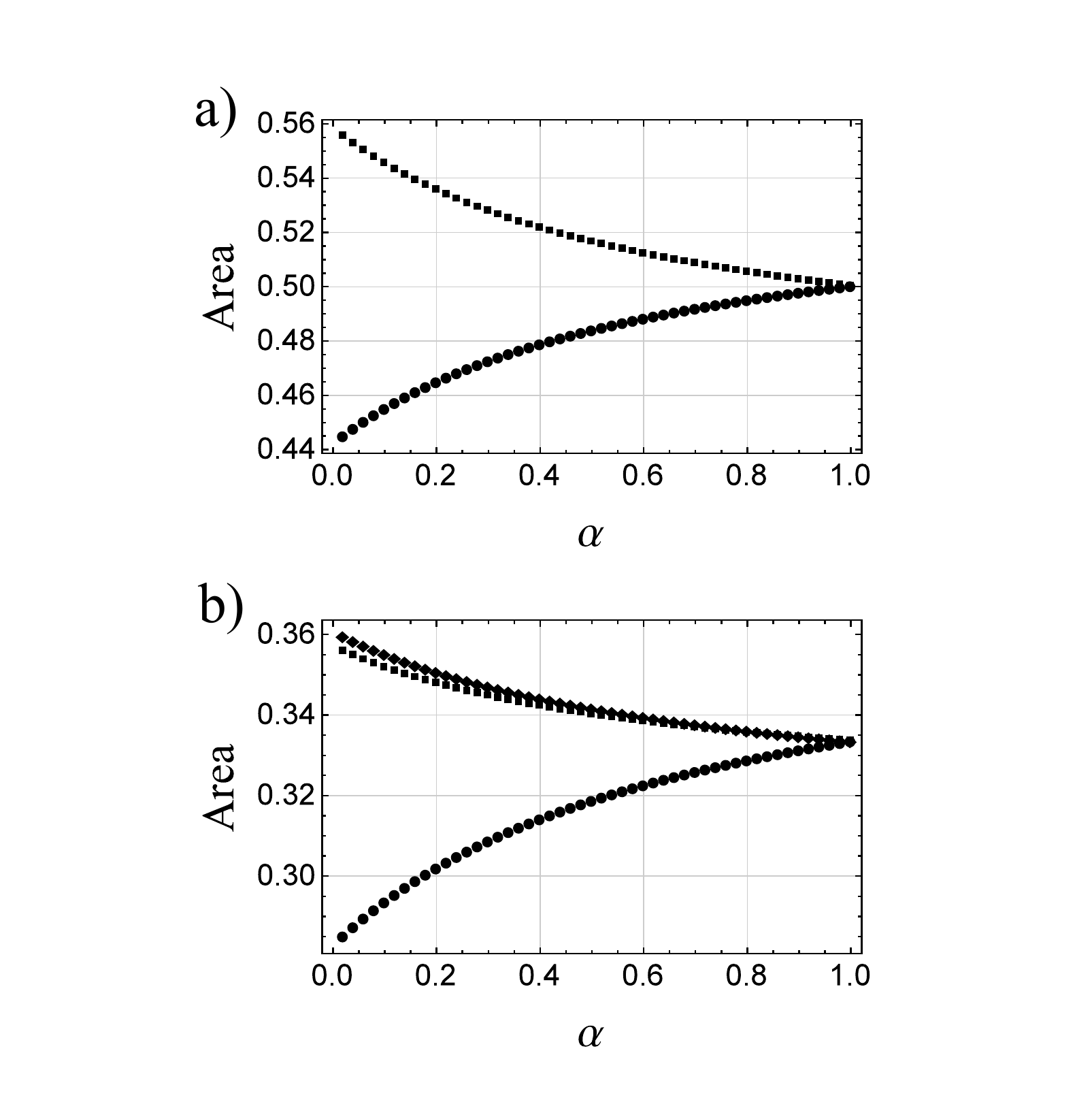}
\caption{ A comparison of the integral of expression (\ref{areaint}%
) for\textbf{\ (a)} $\mathbb{J}_{2}^{(\alpha )}$ and \textbf{(b)} $\mathbb{J}%
_{3}^{(\alpha )}.$ The ($\bullet $) represents $\int_{0}^{\mathbb{N}%
_{n}^{(\alpha )}(1)}\left( \mathbb{J}_{n}^{(\alpha )}\right) ^{2}dx,$ ($%
\blacksquare $) represents $\int_{\mathbb{N}_{n}^{(\alpha )}(1)}^{\mathbb{N}%
_{n}^{(\alpha )}(2)}\left( \mathbb{J}_{n}^{(\alpha )}\right) ^{2}dx,$ ($%
\blacklozenge $) represents $\int_{\mathbb{N}_{n}^{(\alpha )}(1)}^{1}\left( 
\mathbb{J}_{n}^{(\alpha )}\right) ^{2}dx.$ The relative area on $\left( 
\mathbb{J}_{n}^{(\alpha )}\right) ^{2}$ between 0 and $\mathbb{N}%
_{n}^{(\alpha )}(1)$ decreases whereas the relative area between all higher
zeros increases.}
\end{figure}

\begin{figure}
\includegraphics[trim=75  110 75 75,clip, width=\textwidth]{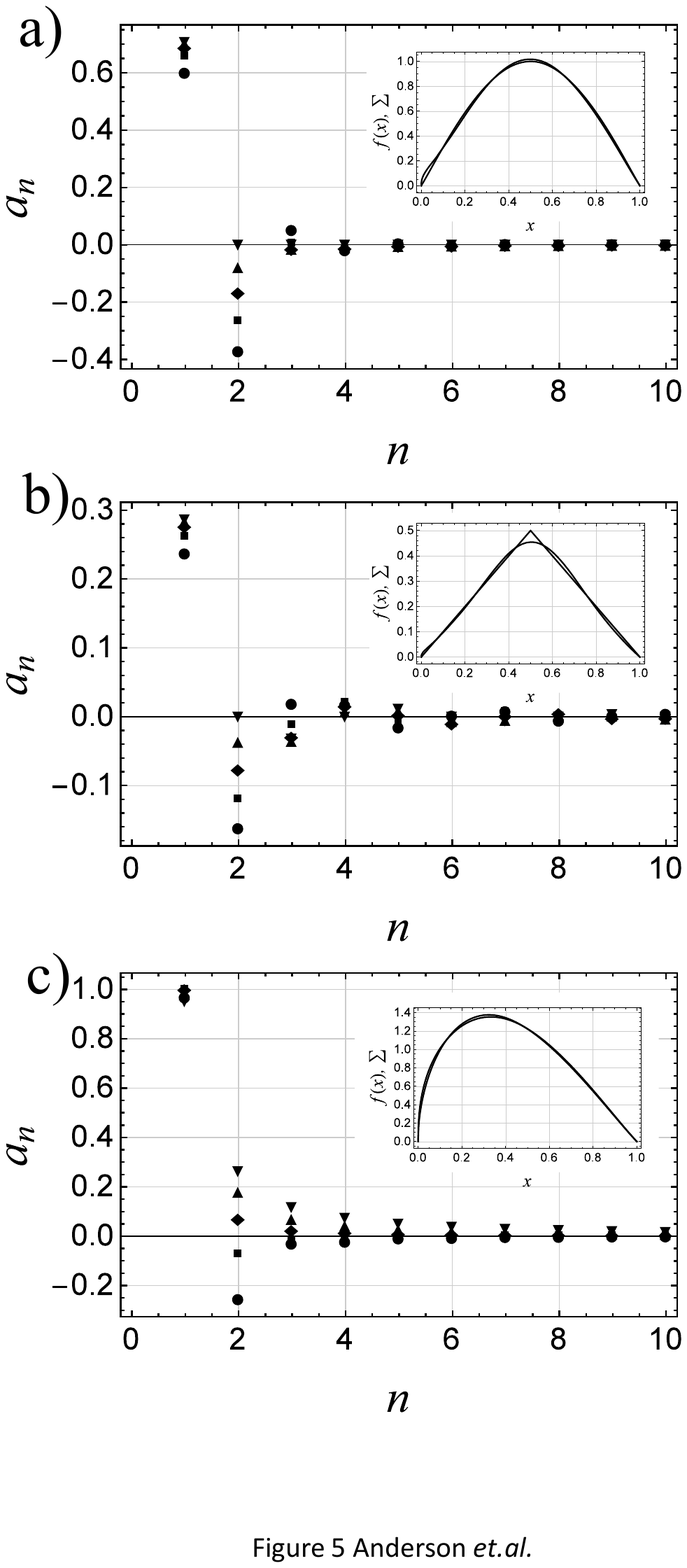}
\caption{  Generalized harmonic spectral decomposition series for
representative functions, \textbf{(a)}\ $f(x)=\sin \pi x,$ \textbf{(b)} $%
f(x)=\frac{1}{2}-\left\vert x-\frac{1}{2}\right\vert $ and \textbf{(c)} $%
f(x)=\mathbb{J}_{1}^{(1/2)}(x).$ $\alpha =$ 1/5 ($\bullet $), 2/5 ($%
\blacksquare $), 3/5 ($\blacklozenge $), 4/5 ($\blacktriangle $), 1 ($%
\blacktriangledown $). The insets show the representative function and a
truncated series. For (a) and (c) one 2 terms in the series were used while
5 terms were used for (b). If more terms are used the partial sum function
becomes visually identical to the representative function (except for near
the sharp point in (b)).}
\end{figure}

One interesting integral to consider is 
\begin{equation}
\int_{\mathbb{N}_{n}^{(\alpha )}(k)}^{\mathbb{N}_{n}^{(\alpha )}(k+1)}\left( 
\mathbb{J}_{n}^{(\alpha )}\right) ^{2}dx.  \label{areaint}
\end{equation}%
This is the area between adjacent zeros and is shown in Fig. 4 for $\mathbb{J%
}_{2}^{(\alpha )}$ and $\mathbb{J}_{3}^{(\alpha )}.$ Note that the area
between $x=0$ and $x=$ $\mathbb{N}_{n}^{(\alpha )}(1)$ decreases with
decreasing $\alpha $ whereas the areas between higher zeros increase. The
difference in area for the higher zeros gets smaller (not shown for $n>3$).

\subsection{Relation to the Fourier sine series and the Fourier-Bessel
series.}

Since the $\mathbb{J}_{n}^{(\alpha )}$ functions form a complete orthonormal
set, they can serve as a basis for expansion of arbitrary functions over the
domain $0\leq x\leq 1$ 
\begin{equation}
f(x)=\sum\limits_{n=1}^{\infty }a_{n}\mathbb{J}_{n}^{(\alpha )}(x),
\label{bigJser}
\end{equation}%
where coefficients are obtained in the usual way via 
\begin{equation}
a_{n}=\int_{0}^{1}f(x)\mathbb{J}_{n}^{(\alpha )}(x)dx.
\end{equation}%
Figure 5 shows the spectral decomposition of a few representative examples: 
$f(x)=\sin \pi x,$ $f(x)=\frac{1}{2}-\left\vert x-\frac{1}{2}\right\vert $
(a triangle waveform), and $f(x)=\mathbb{J}_{1}^{(\frac{1}{2})}(x).$ The
first two of these functions are symmetric about $x=\frac{1}{2}$.
Expectedly, their spectral decompositions show a decreasing amount of $a_{1}$
and an increasing amount of $a_{2}$ as $\alpha $ decreases because the basis
functions are becoming more skewed to the left. The insets in the figures
show how a truncated series compares to $f(x).$

One can use the definition of $\mathbb{J}_{n}^{(\alpha )}$ (Eq. (\ref{bigJ}%
)) in Eq. (\ref{bigJser}) to obtain a relation to the Fourier-Bessel series,%
\begin{equation}
f(x)=\sum\limits_{n=1}^{\infty }a_{n}N_{n}\sqrt{x^{\alpha }}J_{\eta }\left(
n_{\eta }(x^{\alpha })^{\frac{1}{2\eta }}\right),
\end{equation}%
where $N_{n}=1/\sqrt{(\eta -1)J_{\eta -1}(n_{\eta })J_{\eta +1}(n_{\eta })}.$
Letting $c_{n}\equiv a_{n}N_{n}$ and $z=(x^{\alpha })^{\frac{1}{2\eta }}=x^{%
\frac{1+\alpha }{2}}$ yields 
\begin{equation}
f(z)=z^{\eta }\sum\limits_{n=1}^{\infty }c_{n}J_{\eta }\left( n_{\eta
}z\right),
\end{equation}%
where the summation factor is recognized as the well-known Fourier-Bessel
series \cite{Papoulis}. Expressing 
\begin{equation}
\frac{f(z)}{z^{\eta }}=g(z)=\sum\limits_{n=1}^{\infty }c_{n}J_{\eta }\left(
n_{\eta }z\right)
\end{equation}%
where%
\begin{equation}
c_{n}=\int_{0}^{1}zg(z)J_{\eta }\left( n_{\eta }z\right) dz.
\end{equation}%
Often the $c_{n}$ needs to be calculated numerically but several important
cases do yield analytic representations. First, consider the case when $%
g(z)=1.$ In this situation%
\begin{eqnarray}
c_{n} &=&\int_{0}^{1}zJ_{\eta }\left( n_{\eta }z\right) dz  \notag \\
&=&\frac{n_{\eta }^{\eta }\,_{1}F_{2}\left( \frac{\eta }{2}+1;\frac{\eta }{2}%
+2,\eta +1;-\frac{1}{4}n_{\eta }^{2}\right) }{2^{\eta }\left( \eta +2\right)
\Gamma (\eta +1)},
\end{eqnarray}%
where $_{1}F_{2}$ is the hypergeometric function and $\Gamma $ is the gamma
function. Substitution back for $x$ and $a_{n}$ gives. 
\begin{equation}
\sqrt{x^{\alpha }}=\sum\limits_{n=1}^{\infty }\frac{c_{n}}{N_{n}}\mathbb{J}%
_{n}^{(\alpha )}(x).
\end{equation}

More generally consider the case when $g(z)=z^{\gamma -1}$ where $\gamma >-1$. Then,%
\begin{eqnarray}
c_{n} &=&\int_{0}^{1}z^{\gamma }J_{\eta }\left( n_{\eta }z\right) dz \\
&=&\frac{n_{\eta }^{\eta }\,_{1}F_{2}\left( \frac{1}{2}\left( \gamma +\eta
+1\right) ;\frac{1}{2}\left( \gamma +\eta +3\right) ,\eta +1;-\frac{1}{4}%
n_{\eta }^{2}\right) }{2^{\eta }\Gamma \left( \frac{1}{2}\left( \gamma +\eta
+2\right) \right) \Gamma (\eta +1)}.  \notag
\end{eqnarray}%
The particular values of $\gamma =m+1-\frac{\alpha }{2},$ where $m$ is an
integer, gives the monomials, $f(x)=x^{m}$.

\subsection{Relation of $\mathbb{J}_{n}^{(\protect\alpha )}$ to the
confluent hypergeometric functions}

It is well-known that the Bessel functions are related to the confluent
hypergeometric (or Kummer) functions $_{1}F_{1}(a;b;x)$ \cite{AS}. It turns
out that expressing $\mathbb{J}_{n}^{(\alpha )}$ in terms of $%
_{1}F_{1}(a;b;x)$ can be done but leads to a fairly complicated function. It
is perhaps better to use the confluent hypergeometric limit function\cite%
{Petk,MW1} $_{0}F_{1}(;b;x)$ relation to the Bessel functions. One can
employ 
\begin{equation}
J_{\eta }(z)=\frac{z^{2}\,_{0}F_{1}\left( ;\eta +1;-\frac{1}{4}z^{2}\right) 
}{2^{2}\Gamma (\eta +1)}.
\end{equation}%
Substitution of this into Eq. (\ref{bigJ}) expresses $\mathbb{J}%
_{n}^{(\alpha )}$ in terms of $_{0}F_{1}$ as 
\begin{equation}
\mathbb{J}_{n}^{(\alpha )}=\frac{n_{\eta }^{\eta }x^{\alpha
}\,_{0}F_{1}\left( ;\eta +1;-\frac{n_{\eta }^{2}}{4}x^{\alpha +1}\right) }{%
2^{2}\Gamma (\eta +1)\sqrt{(\eta -1)J_{\eta -1}(n_{\eta })J_{\eta
+1}(n_{\eta })}}.  \label{bigJhyp}
\end{equation}

\subsection{Conformable Sturm-Liouville systems}

Consider the operator%
\begin{equation}
\hat{S}=\frac{d}{dx}f(x)\frac{d}{dx},
\end{equation}%
which can be expanded as 
\begin{equation}
\hat{S}=f(x)\frac{d^{2}}{dx^{2}}+f^{\prime }(x)\frac{d}{dx}.
\end{equation}%
Now consider the conformable version of $\hat{S},$%
\begin{equation}
\hat{S}_{\alpha /\beta }=D^{\beta }f(x)D^{\alpha }.
\end{equation}%
The question is how $\hat{S}_{\alpha /\beta }$ relates to $\hat{S}.$
Expanding with $D^{\alpha }=x^{1-\alpha }D^{1}$ we see,%
\begin{eqnarray}
\hat{S}_{\alpha /\beta } &=&x^{1-\beta }D^{1}\left[ fx^{1-\alpha }D^{1}%
\right] \\
&=&x^{1-\beta }\left( \left( x^{1-\alpha }fD^{2}+((1-\alpha )x^{-\alpha
}f+x^{1-\alpha }f^{\prime }\right) D^{1}\right)  \notag \\
&=&x^{2-\alpha -\beta }fD^{2}+((1-\alpha )x^{1-\alpha -\beta }f+x^{2-\alpha
-\beta }f^{\prime })D^{1}.  \notag
\end{eqnarray}%
This operator is not in Sturm-Liouville form but can be made so. Define,%
\begin{eqnarray}
h &=&\frac{1}{x^{2-\alpha -\beta }f}\exp \left[ \int^{x}\frac{(1-\alpha
)u^{1-\alpha -\beta }f+u^{2-\alpha -\beta }f^{\prime }}{u^{2-\alpha -\beta }f%
}du\right]  \notag \\
&=&\frac{1}{x^{2-\alpha -\beta }f}\exp \left[ \int^{x}\frac{(1-\alpha )}{u}du%
\right] \exp \left[ \int^{x}\frac{f^{\prime }}{f}du\right]  \notag \\
&=&\frac{1}{x^{2-\alpha -\beta }f}x^{1-\alpha }f  \notag \\
&=&x^{\beta -1}.
\end{eqnarray}%
So,%
\begin{eqnarray}
\widehat{\mathbb{S}}_{\alpha /\beta } &=&h\hat{S}_{\alpha /\beta }  \notag \\
&=&x^{1-\alpha }fD^{2}+((1-\alpha )x^{-\alpha }f+x^{1-\alpha }f^{\prime
})D^{1}  \notag \\
&=&\frac{d}{dx}x^{1-\alpha }f\frac{d}{dx}.
\end{eqnarray}%
Thus a self-adjoint conformable Sturm-Liouville operator is of the form%
\begin{equation}
\widehat{\mathbb{S}}_{\alpha /\beta }=x^{\beta -1}D^{\beta }f(x)D^{\alpha }.
\end{equation}

\noindent\textbf{Examples}

\underline{Case 1}: $\beta =\alpha $ and $f(x)=1.$ Then, 
\begin{equation}
\widehat{\mathbb{S}}_{\alpha /\alpha }=x^{\alpha -1}D^{\alpha }D^{\alpha }=%
\hat{A}_{2\alpha }
\end{equation}%
where is $\hat{A}_{2\alpha }$ is from Eq. (\ref{selfadjAop}).

\underline{Case 2}: $\beta =\alpha $ and $f(x)=x^{n},$ $n\in \mathbb{Z}^{+}$
(the positive integers). Then, 
\begin{equation}
\widehat{\mathbb{S}}_{\alpha /\alpha }=\frac{d}{dx}x^{n+1-\alpha }\frac{d}{dx%
}.
\end{equation}%
Some differential equations and their solutions are 
\begin{equation}
\widehat{\mathbb{S}}_{\alpha /\alpha }y=0
\end{equation}%
giving%
\begin{equation}
y=A\frac{x^{\alpha -n}}{\alpha -n}+B,
\end{equation}%
and
\begin{equation}
\widehat{\mathbb{S}}_{\alpha /\alpha }y=\Lambda
\end{equation}%
giving%
\begin{equation}
y=\frac{\Lambda x^{\alpha -n+1}}{\alpha -n+1}+A\frac{x^{\alpha -n}}{\alpha -n}%
+B.
\end{equation}%
When $n\rightarrow 0$ in each of these equations we get the results from
acting with $\hat{A}_{2\alpha }.$ This eigenvalue equation,%
\begin{equation}
\widehat{\mathbb{S}}_{\alpha /\alpha }y=\Lambda y
\end{equation}%
does not solve except for $n=1$ which gives 
\begin{eqnarray}
y &=&A\left( x^{\alpha }\right) ^{\frac{\alpha -1}{2\alpha }}J_{\frac{\alpha
-1}{\alpha }}\left( \frac{2\sqrt{\Lambda x^{\alpha }}}{\alpha }\right) \\
&&+B\left( x^{\alpha }\right) ^{\frac{\alpha -1}{2\alpha }}J_{\frac{1-\alpha 
}{\alpha }}\left( \frac{2\sqrt{\Lambda x^{\alpha }}}{\alpha }\right) . 
\notag
\end{eqnarray}

\underline{Case 3:} $\beta =\alpha $ and $f(x)=x^{\alpha }$. Then%
\begin{equation}
\widehat{\mathbb{S}}_{\alpha /\alpha }=\frac{d}{dx}x\frac{d}{dx}.
\end{equation}%
Now, 
\begin{equation}
\widehat{\mathbb{S}}_{\alpha /\alpha }y=\Lambda y
\end{equation}%
is simply 
\begin{equation}
y=AJ_{0}\left( 2\sqrt{\Lambda x}\right) +BY_{0}\left( 2\sqrt{\Lambda x}%
\right).
\end{equation}

\underline{Case 4}: $\beta =\alpha $ and $f(x)=x^{p}$ where $p$ is a
rational fraction. \textsc{Mathematica} can not solve this case generally but, a solution
for 
\begin{equation}
\widehat{\mathbb{S}}_{\alpha /\alpha }y=\Lambda y
\end{equation}%
can be discerned to be 
\begin{equation}
y=A(x^{\alpha })^{\left( \frac{\alpha -p}{2\alpha }\right) }J_{\frac{%
p-\alpha }{\kappa }}\left( \frac{2\sqrt{\Lambda }}{\kappa }(x^{\alpha
})^{\left( \frac{\kappa }{2\alpha }\right) }\right) +B(x^{\alpha })^{\left( 
\frac{\alpha -p}{2\alpha }\right) }J_{\frac{a-p}{\kappa }}\left( \frac{2%
\sqrt{\Lambda }}{\kappa }(x^{\alpha })^{\left( \frac{\kappa }{2\alpha }%
\right) }\right) ,
\end{equation}%
where $\kappa =1+\alpha -p.$ This does simplify further as 
\begin{equation}
y=Ax^{\frac{\alpha -p}{2}}J_{\frac{p-\alpha }{\kappa }}\left( \frac{2\sqrt{%
\Lambda }}{\kappa }x^{\frac{\kappa }{2\alpha }}\right) +BAx^{\frac{\alpha -p%
}{2}}J_{\frac{\alpha -p}{\kappa }}\left( \frac{2\sqrt{\Lambda }}{\kappa }x^{%
\frac{\kappa }{2\alpha }}\right). 
\end{equation}%
For the case of boundary conditions at $y(0)=y(1)=0$, 
\begin{equation}
y=Bx^{\frac{\alpha -p}{2}}J_{\frac{\alpha -p}{\kappa }}\left( n_{\frac{%
\alpha -p}{\kappa }}x^{\frac{\kappa }{2\alpha }}\right),   \label{soleq73}
\end{equation}%
where $n_{\frac{p-\alpha }{\kappa }}$ is the $n^{th}$ zero of $J_{\frac{%
\alpha -p}{\kappa }}.$ The normalization constant is%
\begin{equation}
B=\frac{1}{\sqrt{\left( \frac{-1}{1+\alpha -p}\right) J_{\frac{\alpha -p}{%
\kappa }-1}\left( n_{\frac{\alpha -p}{\kappa }}\right) J_{\frac{\alpha -p}{%
\kappa }+1}\left( n_{\frac{\alpha -p}{\kappa }}\right) }}.  \label{soleq74}
\end{equation}%
Figure 6 shows $y$ (left column) and $y^{2}$ (right column) for the case of $%
\alpha =3/4$ and $p=0,$ 1/8, 1/4, 3/8, 1/2, 5/8 for the first three
eigenstates. As $p$ approaches $\alpha $ the graphs compress towards $x=0.$
Figure 7 shows the case where $p=1/4$ and $\alpha =3/8,$ 1/2, 5/8, 3/4, 7/8
and 1 for $y$ for the first and second eigenfunctions.

\begin{figure}
\includegraphics[trim=75  325 75 75,clip, width=\textwidth]{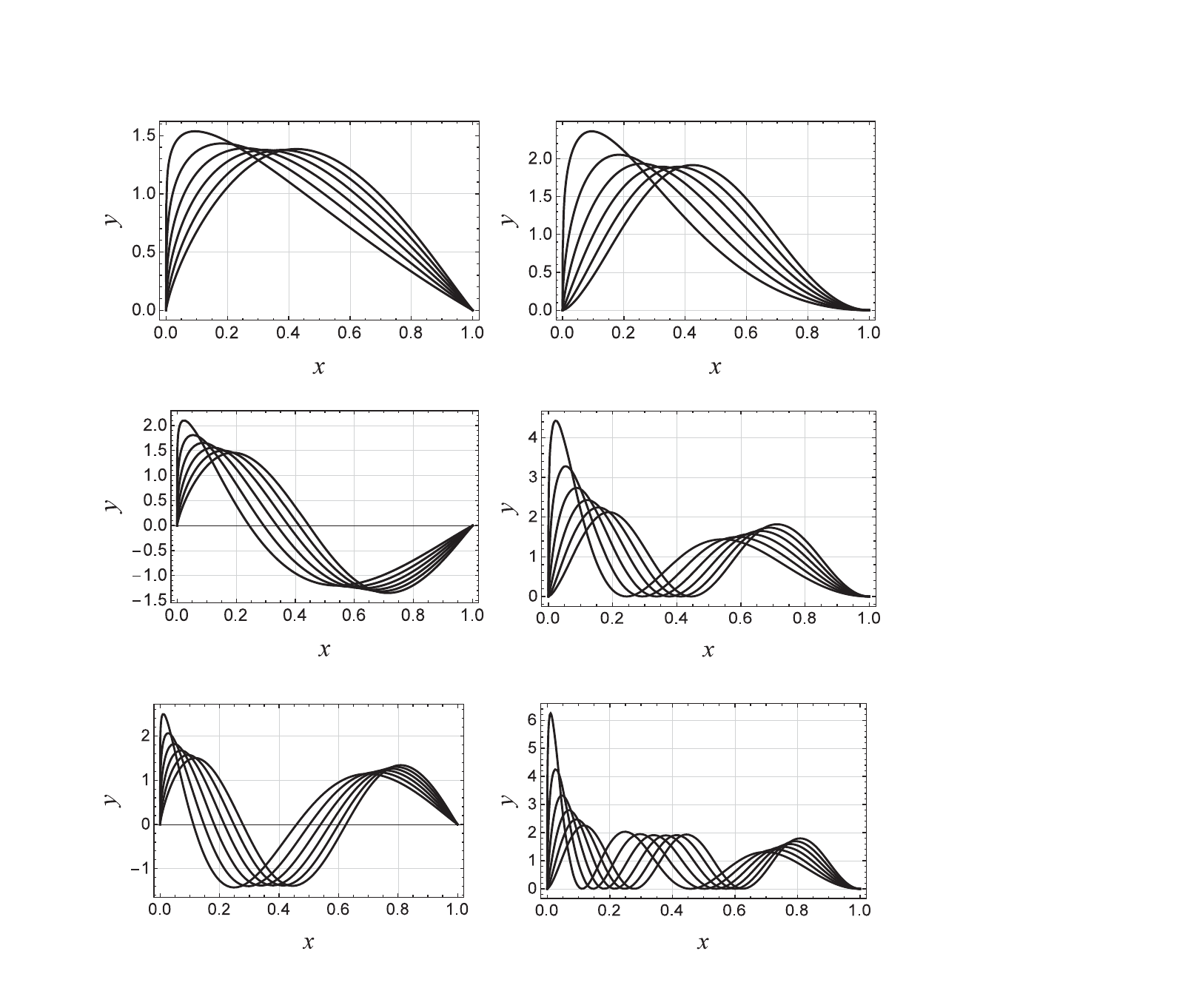}
\caption{ $y$ (left column) and $y^{2}$ (right column) from Eq. (%
\ref{soleq73}) for the case of $\alpha =3/4$ and $p=0,$ 1/8, 1/4, 3/8, 1/2,
5/8 for the first three eigenstates. As $p$ approaches $\alpha $ the graphs
compress towards $x=0.$}
\end{figure}

\begin{figure}
\includegraphics[trim=75  550 75 100,clip, width=\textwidth]{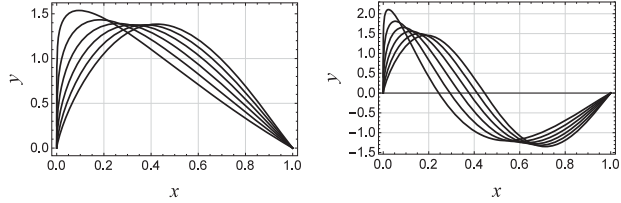}
\caption{The case where $p=1/4$ and $\alpha =3/8,$ 1/2, 5/8, 3/4,
7/8 and 1 for $y$ for the first and second eigenfunctions for Eq. (\ref%
{soleq73}).}
\end{figure}

Case 4 brings up an interesting characteristic in that the solutions depend
only on $\alpha -p\equiv \lambda $ thus what matters for the shape of the
curve of $y$ is how far $p$ is from $\alpha $ not the absolute values of
either. Stepping back further; if $f(x)=x^{p}g(x)$ then 
\begin{eqnarray}
\frac{d}{dx}x^{1-\alpha }f\frac{d}{dx} &=&\frac{d}{dx}x^{1-\alpha }x^{p}g(x)%
\frac{d}{dx}  \notag \\
&=&\frac{d}{dx}x^{1-\alpha +p}g(x)\frac{d}{dx}  \notag \\
&=&\frac{d}{dx}x^{1-\lambda }g(x)\frac{d}{dx}x^{\beta -1}D^{\beta
}g(x)D^{\lambda }.
\end{eqnarray}%
Here $p$ is not restricted to being a rational fraction nor less than $%
\alpha $ but $\lambda $ must still be less than 1.

\underline{Case 5}: Now consider the case when 
\begin{eqnarray}
x^{1-\alpha }f^{\prime }+(1-\alpha )x^{-\alpha }f &=&0  \notag \\
f^{\prime }+(1-\alpha )x^{-1}f &=&0
\end{eqnarray}%
this occurs when $f=x^{\alpha -1}.$ So, consider $\beta =\alpha $ and $%
f(x)=x^{\alpha -1}.$ This solves the eigenvalue equation with the form 
\begin{equation}
y=A\cos \left( \sqrt{\Lambda }x\right) +B\sin \left( \sqrt{\Lambda }x\right).
\end{equation}%
Taken together cases 1--5 suggest the conjecture that Eq. (\ref{soleq73})
and (\ref{soleq74}) is the solution even when $p$ is expanded from a
rational fraction to the reals in which $0<p\leq \alpha .$ And perhaps even
when $0<p\leq \infty .$

Often times one encounters operators of the form

\begin{equation}
\hat{S}=\frac{1}{f(x)}\frac{d}{dx}f(x)\frac{d}{dx},
\end{equation}%
which, when made conformable, becomes
\begin{equation}
\frac{1}{f(x)}\widehat{\mathbb{S}}_{\alpha /\beta }.
\end{equation}
We can consider a special case of this type of operator.

\underline{Case 6}: $f=x^{r}$ where $r$ is any real number greater than
zero. The solutions to 
\begin{equation}
\frac{1}{x^{r}}\widehat{\mathbb{S}}_{\alpha /\beta }y=\Lambda y
\end{equation}%
are 
\begin{equation}
y=A(x^{\alpha })^{\frac{\alpha -r}{2\alpha }}J_{\frac{\alpha -r}{1+a}}\left( 
\frac{2\sqrt{\Lambda }(x^{\alpha })^{\frac{\alpha +1}{2\alpha }}}{1+\alpha }%
\right) +B(x^{\alpha })^{\frac{\alpha -r}{2\alpha }}J_{\frac{r-\alpha }{%
1+\alpha }}\left( \frac{2\sqrt{\Lambda }(x^{\alpha })^{\frac{\alpha +1}{%
2\alpha }}}{1+\alpha }\right).
\end{equation}

\section{Integral transforms}

Any new definition of a fractional/conformable derivative leads naturally to
the consideration of fractional/conformable differential equations and,
subsequently, the use of fractional/conformable Laplace transforms to solve
them. Indeed, numerous versions of fractional/conformable Laplace transforms
have appeared in the literature. Some of these look very much like a regular
Laplace transform \cite{Abdel,Kexue,Jumarie,Salah}, while others look quite
different \cite{Sharma,Deshmukh,Treumann}. The $k$-Laplace transforms \cite%
{Treumann} look a bit more like Mellin transforms, while the definitions
used by Sharma \cite{Sharma}, Deshmukh and Gudadhe \cite{Deshmukh}, and
Gorty \cite{Gorty} involve cotangents and cosecants in the exponential
Laplace kernel. To be sure, the regular Laplace transform has also been used
to tackle fractional differential equations, often resulting in a
Mittag-Leffler expansion solution \cite{Achar}.

In this work, the choice was made to actually explore a conformable
formulation of a Fourier transform, whose conventional counterpart is 
\begin{equation}
\mathfrak{F}[f(t)]=\int_{-\infty }^{\infty }f(t)e^{i\omega t}dt.
\label{FLTreg}
\end{equation}
This is trivially related to the Laplace transform with $s=-i\omega $ and
limiting the integration to $t\geq 0.$ The reason for working with the
Fourier transform is twofold. First, the inverse transform involves
integration on the real axis of the transform variable rather than along the
Bromwich contour in the complex $s$-plane as is done with the inverse
Laplace transform. Second, the Fourier transform treats the forward and
inverse transforms more symmetrically, and with potential applications to
quantum mechanics in mind, this serves as a more natural path. The
conformable Laplace transform is explicitly given at the end of subsection 4.1.5.

\subsection{The conformable Fourier transform}

The Fourier transform is developed here in the context of the definition of
the conformable derivative given in Eq. (\ref{KKder}), the $D^{\alpha }$
operator given in Eq. (\ref{KKderop}) and its corresponding inverse,%
\begin{equation}
I^{\alpha }=\int^{t}\frac{(\cdot )}{\tau ^{1-\alpha }}d\tau .  \label{Iop}
\end{equation}%
That is, for $\tau =(\alpha x)^{\frac{1}{\alpha }}\implies d\tau =(\alpha
x)^{\frac{1}{\alpha }-1}dx$ and endpoints $t_{0}\mapsto (\alpha t_{0})^{%
\frac{1}{\alpha }}=t_{0}^{\prime }$, $t\mapsto (\alpha t)^{\frac{1}{\alpha }%
}=t^{\prime }$, we have

\begin{equation}
I^{\alpha }\left[ f(t)\right] =\int_{t_{0}}^{t}f(\tau )\tau ^{\alpha
-1}d\tau =\int_{t_{0}^{\prime }}^{t^{\prime }}f\left[ (\alpha x)^{\frac{1}{%
\alpha }}\right] dx.  \label{Iop2}
\end{equation}%
Abdeljawad has recently explored a Laplace transform in this context.\cite%
{Abdel} (Note: similar to the change of variables technique used in creating
conformable differential equations from ordinary differential equations, one
may apply Eq (\ref{Iop2}) to create a conformable integral equation from an
ordinary integral equation.)

In the reverse direction, if one has any integral in the form $%
\int_{t_{0}^{\prime }}^{t^{\prime }}f(x)dx$ one may create a conformable
integral from it via $x=\frac{\tau ^{\alpha }}{\alpha }\implies dx=\tau
^{\alpha -1}d\tau $ and endpoints $t_{0}=\frac{(t_{0}^{\prime })^{\alpha }}{%
\alpha }$ and $t=\frac{(t^{\prime })^{\alpha }}{\alpha }$. That is

\begin{equation}
\int_{t_{0}^{\prime }}^{t^{\prime }}f(x)dx=\int_{t_{0}}^{t}f\left( \frac{%
\tau ^{\alpha }}{\alpha }\right) \tau ^{\alpha -1}d\tau =I^{\alpha }\left[
g(t)\right] 
\end{equation}%
for $g(\tau )=f\left( \frac{\tau ^{\alpha }}{\alpha }\right) $.

We do precisely this to recover a conformable Fourier transform. That is,

\begin{eqnarray}
\mathbb{F}\left[ f\left[ \left( \alpha t\right) ^{\frac{1}{\alpha }}\right] %
\right]  &=&\int_{-\infty }^{\infty }f\left[ \left( \alpha t\right) ^{\frac{1%
}{\alpha }}\right] e^{-ist}dt \notag \\
&=&\int_{-\infty }^{0}f\left[ \left( \alpha
t\right) ^{\frac{1}{\alpha }}\right] e^{-ist}dt+\int_{0}^{\infty }f\left[
\left( \alpha t\right) ^{\frac{1}{\alpha }}\right] e^{-ist}dt  \notag \\
&=&\int_{\Gamma ^{\alpha }}^{0}f(\tau )e^{-is\frac{\tau ^{a}}{\alpha }}\tau
^{\alpha -1}d\tau +\int_{0}^{\infty }f(\tau )e^{-is\frac{\tau ^{a}}{\alpha }%
}\tau ^{\alpha -1}d\tau   \notag \\
&=&\mathbb{F}_{\alpha }\left[ f(\tau )\right] \tilde{f}_{\alpha }(s),
\end{eqnarray}%
where the notation $\int_{\Gamma ^{\alpha }}^{0}$ indicates integration
along the complex ray $re^{i\pi /\alpha }$ where $r\in (0,\infty )$ (see Fig. 8). In this
manner, one may compute a conformable Fourier transform from a special case
of the regular Fourier transform. 

\begin{figure}
\includegraphics[trim=75  250 75 75,clip, width=\textwidth]{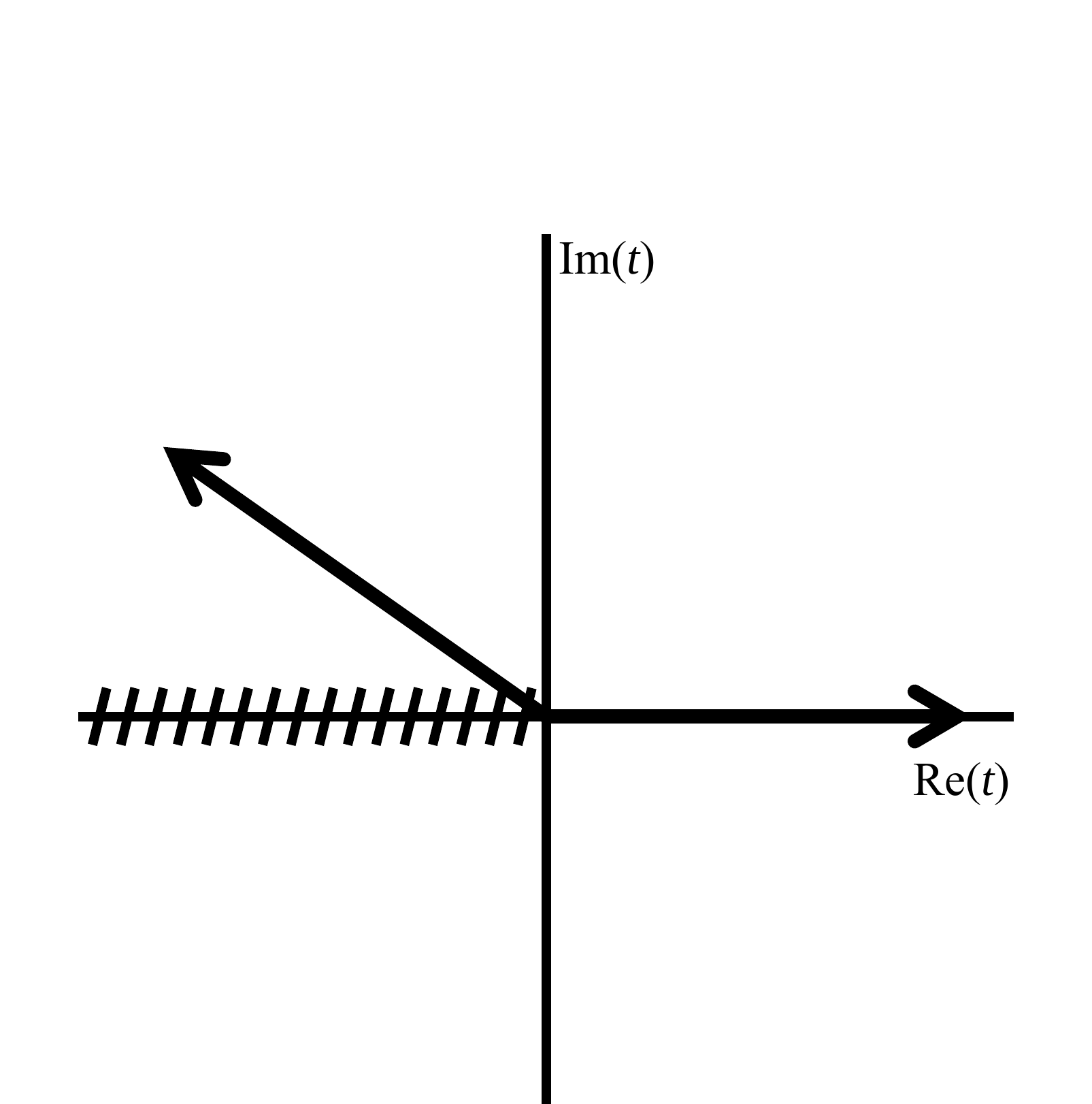}
\caption{The integration contour for the conformable Fourier
transform pair (Eqs. (\ref{fracFLTdef}) and (\ref{fractFLTinvdef})).}
\end{figure}

The conformable Fourier transform carries with it some difficulties when
viewed in the context of the change of variable because as $t\rightarrow
t^{\alpha }/\alpha $ one no longer can integrate over the negative real
values of $t.$ However the same change of variable suggests a suitable
integration contour as depicted in Fig. 8. When viewed in the complex plane,
one can avoid the branch cut created along the negative real axis by
bending the integration contour in Eq. (\ref{FLTreg}) to that along the ray
formed by $re^{\frac{i\pi }{\alpha }}.$ The symbol $\int_{\Gamma ^{\alpha
}}^{\infty }$ is used to represent integration along the contour shown in
Fig. 8.

Looking at the inverse transform, the same pattern emerges, giving%
\begin{eqnarray}
2\pi \mathbb{F}^{-1}\left[ \tilde{f}_{\alpha }\left[ \left( \beta s\right) ^{%
\frac{1}{\alpha }}\right] \right] &=&\int_{-\infty }^{\infty }\tilde{f}%
_{\alpha }\left[ \left( \beta s\right) ^{\frac{1}{\alpha }}\right] e^{-ist}ds
\notag \\
&=&\int_{-\infty }^{0}\tilde{f}_{\alpha }\left[ \left( \beta s\right) ^{%
\frac{1}{\alpha }}\right] e^{-ist}ds+\int_{0}^{\infty }\tilde{f}_{\alpha }%
\left[ \left( \beta s\right) ^{\frac{1}{\alpha }}\right] e^{-ist}ds  \notag
\\
&=&\int_{\Gamma ^{\beta }}^{0}\tilde{f}_{\alpha }\left( \omega \right) e^{-i%
\frac{\omega ^{\beta }}{\beta }t}\omega ^{\beta -1}d\omega +\int_{0}^{\infty
}\tilde{f}_{\alpha }\left( \omega \right) e^{-i\frac{\omega ^{\beta }}{\beta 
}t}\omega ^{\beta -1}d\omega  \notag \\
&=&2\pi \mathbb{F}^{-1}\left[ \tilde{f}_{\alpha }\left( \omega \right) %
\right] =2\pi f_{\alpha /\beta }(t).
\end{eqnarray}
Thus the conformable Fourier transform pair is%
\begin{eqnarray}
\mathbb{F}_{\alpha /\beta }[f(t)] &=&\tilde{f}_{\alpha /\beta }(\omega )
\label{fracFLTdef} \\
&=&\int_{\Gamma ^{\alpha }}^{\infty }f(t)e^{\frac{i}{\alpha \beta }\omega
^{\beta }t^{\alpha }}dt^{\alpha }  \notag \\
&=&\int_{0}^{\infty }f\left( \frac{t^{\alpha }}{\alpha }\right) e^{\frac{i}{%
\alpha \beta }\omega ^{\beta }t^{\alpha }}t^{\alpha -1}dt+\int_{0}^{\infty
}f\left( -\frac{t^{\alpha }}{\alpha }\right) e^{\frac{-i}{\alpha \beta }%
\omega ^{\beta }t^{\alpha }}t^{\alpha -1}dt  \notag
\end{eqnarray}%
and 
\begin{equation}
\mathbb{F}_{\alpha /\beta }^{-1}[\tilde{f}_{a/\beta }(\omega )]=f(t)=\frac{1%
}{2\pi }\int_{\Gamma ^{\beta }}^{\infty }\tilde{f}_{\alpha /\beta }(\omega
)e^{-\frac{i}{\alpha \beta }\omega ^{\beta }t^{\alpha }}d\omega ^{\beta },
\label{fractFLTinvdef}
\end{equation}%
where $0<\alpha ,\beta \leq 1$ (the inverse is verified below). This
transformation connects $\frac{t^{\alpha }}{\alpha }$-space to $\frac{\omega
^{\beta }}{\beta }$-space in analogy with $t$-space to $\omega $-space for a
regular Fourier transform. One can now consider a number of properties of
this definition of the conformable Fourier transform including verifying the
transform pair is one-to-one, the transform of the derivative, the
derivative in transform space and the convolution theorem.

Note that in the limit $\alpha \rightarrow 1$ and $\beta \rightarrow 1,$%
\begin{eqnarray}
\lim_{\alpha \rightarrow 1,\beta \rightarrow 1}\mathbb{F}_{\alpha /\beta
}[f(t)] &=&\int_{0}^{\infty }f\left( t\right) e^{i\omega
t}dt+\int_{0}^{\infty }f\left( -t\right) e^{-i\omega t}dt  \notag \\
&=&\int_{-\infty }^{\infty }f\left( t\right) e^{i\omega t}dt=\mathfrak{F}%
[f(t)].
\end{eqnarray}

\subsubsection{Conformable Dirac $\protect\delta $-function}

The Dirac $\delta $-function plays an important role in this analysis and it
is useful to define a fractional version of it. Define the conformable Dirac 
$\delta $-function to be 
\begin{equation}
\delta \left( x^{\alpha }-x_{0}^{\alpha }\right) \equiv \frac{1}{2\pi }%
\int_{\Gamma ^{\beta }}^{\infty }e^{\frac{i}{\alpha \beta }y^{\beta }\left(
x^{\alpha }-x_{0}^{\alpha }\right) }dy^{\beta }.  \label{deldef}
\end{equation}

One can investigate the important case of $\delta \left( f(x^{\alpha
})\right) .$ With the substitution $v=x^{\alpha },$ one is able to employ
the analogous property for the regular $\delta $-function. Namely, 
\begin{equation}
\delta (f(v))=\sum_{\text{roots}}\frac{\delta (v-v_{0})}{\left\vert
f^{\prime }(v_{0})\right\vert },
\end{equation}%
where each root, $v_{0},$ of $f$ provides a term in the summation.
Substituting back for $x$ gives 
\begin{eqnarray}
\delta \left( f(x^{\alpha })\right) &=&\sum_{\text{roots}}\frac{\delta
(x^{\alpha }-v_{0})}{\left\vert f^{\prime }(v_{0})\right\vert }  \notag \\
&=&\sum_{\text{roots}}\frac{\delta (x-v_{0}^{1/\alpha })}{\alpha \left(
v_{0}^{1/\alpha }\right) ^{\alpha -1}\left\vert f^{\prime
}(v_{0})\right\vert }.  \label{delfunct}
\end{eqnarray}

Two important results that will be used subsequently follow. First, 
\begin{equation}
\delta (x^{\alpha }-x_{0}^{\alpha })=\frac{\delta (x-x_{0})}{\alpha
x_{0}^{\alpha -1}}.  \label{deltadef1}
\end{equation}%
Second, $\delta (x^{\alpha }-\left( x_{1}^{\alpha }-x_{2}^{\alpha }\right) )$
can be obtained using Eq. (\ref{delfunct}) to be%
\begin{equation}
\delta (x^{\alpha }-\left( x_{1}^{\alpha }-x_{2}^{\alpha }\right) )=\frac{%
\delta (x-\left( x_{1}^{\alpha }-x_{2}^{\alpha }\right) ^{1/\alpha })}{%
\alpha \left( \left( x_{1}^{\alpha }-x_{2}^{\alpha }\right) ^{1/\alpha
}\right) ^{\alpha -1}}.  \label{deltadef2}
\end{equation}

\subsubsection{Inversion pair}

Verifying the inversion pair%
\begin{eqnarray}
\mathbb{F}_{\alpha /\beta }^{-1}[\mathbb{F}_{\alpha /\beta }[f(t)]] &=&\frac{%
1}{2\pi }\int_{\Gamma ^{\beta }}^{\infty }\int_{\Gamma _{1}^{\alpha
}}^{\infty }f(t_{1})e^{\frac{i}{\alpha \beta }\omega ^{\beta }t^{\alpha
}}t_{1}^{\alpha -1}dt_{1}e^{-\frac{i}{\alpha \beta }\omega ^{\beta
}t^{\alpha }}d\omega ^{\beta }  \notag \\
&=&\int_{\Gamma _{1}^{\alpha }}^{\infty }f(t_{1})t_{1}^{\alpha -1}\frac{1}{%
2\pi }\int_{\Gamma ^{\beta }}^{\infty }e^{\frac{i}{\alpha \beta }\omega
^{\beta }\left( t_{1}^{\alpha }-t^{\alpha }\right) }d\omega ^{\beta }dt_{1}.
\notag \\
&=&\alpha \int_{\Gamma _{1}^{\alpha }}^{\infty }f(t_{1})t_{1}^{\alpha
-1}\delta (t_{1}^{\alpha }-t^{\alpha })dt_{1}.  \label{ForwardA}
\end{eqnarray}%
Using Eq. (\ref{deltadef1}) 
\begin{eqnarray}
\mathbb{F}_{\alpha /\beta }^{-1}[\mathbb{F}_{\alpha /\beta }[f(t)]]
&=&\alpha \int_{\Gamma _{1}^{\alpha }}^{\infty }f(t_{1})t_{1}^{\alpha -1}%
\frac{\delta (t_{1}-t)}{\alpha t_{1}^{\alpha -1}}dt_{1}  \notag \\
&=&\int_{-\infty }^{\infty }f(t_{1})\delta (t_{1}-t)dt_{1}  \notag \\
&=&f(t).
\end{eqnarray}%
Indeed one recovers the original function. For completeness one can also
verify that 
\begin{equation}
\mathbb{F}_{\alpha /\beta }[\mathbb{F}_{\alpha /\beta }^{-1}[\tilde{f}%
_{\alpha /\beta }(\omega )]]=\tilde{f}_{a/\beta }(\omega )  \label{reverse}
\end{equation}%
in a similar way.

\subsubsection{The derivative and the transform-space derivative}

Consider the conformable Fourier transform of the conformable derivative 
\begin{eqnarray}
\mathbb{F}_{\alpha /\beta }[D^{\alpha }\left[ f\right] ] &=&\int_{0}^{\infty
}D^{\alpha }\left[ f\right] e^{\frac{i}{\alpha \beta }\omega ^{\beta
}t^{\alpha }}dt^{\alpha }  \notag \\
&=&\int_{0}^{\infty }t^{1-\alpha }\frac{df}{dt}e^{\frac{i}{\alpha \beta }%
\omega ^{\beta }t^{\alpha }}t^{\alpha -1}dt  \notag \\
&=&\int_{0}^{\infty }\frac{df}{dt}e^{\frac{i}{\alpha \beta }\omega ^{\beta
}t^{\alpha }}dt.  \label{DerivativeA}
\end{eqnarray}%
Integration by parts gives 
\begin{eqnarray}
\mathbb{F}_{\alpha /\beta }[D^{\alpha }\left[ f\right] ] &=&-i\frac{\omega
^{\beta }}{\beta }\int_{\Gamma ^{\alpha }}^{\infty }fe^{\frac{i}{\alpha
\beta }\omega ^{\beta }t^{\alpha }}t^{\alpha -1}dt  \notag \\
&=&-i\frac{\omega ^{\beta }}{\beta }\int_{\Gamma ^{\alpha }}^{\infty }fe^{%
\frac{i}{\alpha \beta }\omega ^{\beta }t^{\alpha }}dt^{\alpha }  \notag \\
&=&-i\frac{\omega ^{\beta }}{\beta }\tilde{f}_{\alpha /\beta }(\omega )
\label{DerivativeB}
\end{eqnarray}%
and one recovers the usual formula with $\frac{\omega ^{\beta }}{\beta }$
replacing $\omega $. Consider the case where $1<\kappa \leq 2.$ Now let $%
\alpha =\kappa -1$,%
\begin{eqnarray}
\mathbb{F}_{\alpha /\beta }[D^{\kappa }\left[ f\right] ] &=&\mathbb{F}%
_{\alpha /\beta }[D^{\alpha }D^{1}\left[ f\right] ]  \notag \\
&=&\int_{\Gamma ^{\alpha }}^{\infty }D^{\alpha }f^{\prime }e^{\frac{i}{%
\alpha \beta }\omega ^{\beta }t^{\alpha }}dt^{\alpha },
\end{eqnarray}%
which is just like Eq. (\ref{DerivativeB}) but with $f$ replaced by $%
f^{\prime }.$ Thus 
\begin{equation}
\mathbb{F}_{\alpha /\beta }[D^{\kappa }\left[ f\right] ]=-i\frac{\omega
^{\beta }}{\beta }\int_{\Gamma ^{\alpha }}^{\infty }f^{\prime }e^{\frac{i}{%
\alpha \beta }\omega ^{\beta }t^{\alpha }}dt^{\alpha }.  \label{derform}
\end{equation}%
The second term is just the conformable Fourier transform of the conformable
derivative so, ultimately,%
\begin{equation}
\mathbb{F}_{\alpha /\beta }[D^{\kappa }\left[ f\right] ]=\left( \frac{\omega
^{\beta }}{\beta }\right) ^{2}\tilde{f}_{\alpha /\beta }(\omega ).
\end{equation}%
This same process could be carried out for $\kappa >2.$

Turning now to the conformable Fourier transform of the function $\frac{t^{\alpha }}{%
\alpha }f$ one sees,%
\begin{eqnarray}
\mathbb{F}_{\alpha /\beta }\left[ \frac{t^{\alpha }}{\alpha }f\right]
&=&\int_{\Gamma ^{\alpha }}^{\infty }\frac{t^{\alpha }}{\alpha }fe^{\frac{i}{%
\alpha \beta }\omega ^{\beta }t^{\alpha }}dt^{\alpha }  \notag \\
&=&\int_{\Gamma ^{\alpha }}^{\infty }D_{\omega }^{\alpha }\left[ fe^{\frac{i%
}{\alpha \beta }\omega ^{\beta }t^{\alpha }}\right] dt^{\alpha }  \notag \\
&=&D_{\omega }^{\alpha }\left[ \int_{\Gamma ^{\alpha }}^{\infty }fe^{\frac{i%
}{\alpha \beta }\omega ^{\beta }t^{\alpha }}dt^{\alpha }\right]  \notag \\
&=&D_{\omega }^{\alpha }\left[ \mathbb{F}_{\alpha /\beta }\left[ f\right] %
\right],
\end{eqnarray}%
where $D_{\omega }^{\alpha }$ means the conformable derivative in $\omega $
space.

\subsubsection{Conformable convolution and the conformable convolution
theorem}

Consider the inverse Fourier transform of the product, $\tilde{f}_{\alpha
/\beta }(\omega )\tilde{g}_{\alpha /\beta }(\omega )$. Here,%
\begin{eqnarray}
\mathbb{F}_{\alpha /\beta }^{-1}\left[ \tilde{f}_{\alpha /\beta }\tilde{g}%
_{\alpha /\beta }\right] &=&\frac{1}{2\pi }\int_{\Gamma ^{\beta }}^{\infty
}\int_{\Gamma _{1}^{\alpha }}^{\infty }f(t_{1})e^{\frac{i}{\alpha \beta }%
\omega ^{\beta }t_{1}^{\alpha }}dt_{1}^{\alpha }\int_{\Gamma _{2}^{\alpha
}}^{\infty }g(t_{2})e^{\frac{i}{\alpha \beta }\omega ^{\beta }t_{2}^{\alpha
}}dt_{2}^{\alpha }e^{-\frac{i}{\alpha \beta }\omega ^{\beta }t^{\alpha
}}d\omega ^{\beta }  \notag \\
&=&\int_{\Gamma _{1}^{\alpha }}^{\infty }f(t_{1})\int_{\Gamma _{2}^{\alpha
}}^{\infty }g(t_{2})\frac{1}{2\pi }\int_{\Gamma ^{\beta }}^{\infty }e^{-%
\frac{i}{\alpha \beta }\omega ^{\beta }\left( t^{\alpha }-t_{1}^{\alpha
}-t_{2}^{\alpha }\right) }d\omega ^{\beta }dt_{2}^{\alpha }dt_{1}^{\alpha } 
\notag \\
&=&\alpha \int_{\Gamma _{1}^{\alpha }}^{\infty }f(t_{1})\int_{\Gamma
_{2}^{\alpha }}^{\infty }g(t_{2})\delta \left( t_{2}^{\alpha }+t_{1}^{\alpha
}-t^{\alpha }\right) .
\end{eqnarray}%
Using Eq. (\ref{deltadef2}),%
\begin{eqnarray}
\mathbb{F}_{\alpha /\beta }^{-1}\left[ \tilde{f}_{\alpha /\beta }\tilde{g}%
_{\alpha /\beta }\right] &=&\alpha \int_{\Gamma _{1}^{\alpha }}^{\infty
}f(t_{1})\int_{\Gamma _{2}^{\alpha }}^{\infty }g(t_{2})\frac{\delta
(t_{2}-\left( t^{\alpha }-t_{1}^{\alpha }\right) ^{1/\alpha })}{\alpha
\left( \left( t^{\alpha }-t_{1}^{\alpha }\right) ^{1/\alpha }\right)
^{\alpha -1}}t_{2}^{\alpha -1}dt_{2}dt_{1}^{\alpha }  \notag \\
&=&\int_{\Gamma _{1}^{\alpha }}^{\infty }f(t_{1})g\left( \left( t^{\alpha
}-t_{1}^{\alpha }\right) ^{1/\alpha }\right) \frac{\left( \left( t^{\alpha
}-t_{1}^{\alpha }\right) ^{1/\alpha }\right) ^{\alpha -1}}{\left( \left(
t^{\alpha }-t_{1}^{\alpha }\right) ^{1/\alpha }\right) ^{\alpha -1}}%
dt_{1}^{\alpha }  \notag \\
&=&\int_{\Gamma _{1}^{\alpha }}^{\infty }f(t_{1})g\left( \left( t^{\alpha
}-t_{1}^{\alpha }\right) ^{1/\alpha }\right) dt_{1}^{\alpha }.
\end{eqnarray}%
At this point make a substitution $v=\frac{t_{1}^{\alpha }}{\alpha },$ 
\begin{equation}
\mathbb{F}_{\alpha /\beta }^{-1}\left[ \tilde{f}_{\alpha /\beta }\tilde{g}%
_{\alpha /\beta }\right] =\int_{-\infty }^{\infty }f\left( \left( \alpha
v\right) ^{1/\alpha }\right) g\left( \left( \frac{\alpha t^{\alpha }}{\alpha 
}-\alpha v\right) ^{1/\alpha }\right) dv.
\end{equation}%
For convenience one can recast the functions as 
\begin{equation}
\begin{array}{c}
f(x)=F\left( \frac{x^{\alpha }}{\alpha }\right) \\ 
g(x)=G\left( \frac{x^{\alpha }}{\alpha }\right).%
\end{array}%
\end{equation}%
Doing so gives 
\begin{equation}
\mathbb{F}_{\alpha /\beta }^{-1}\left[ \tilde{f}_{\alpha /\beta }\tilde{g}%
_{\alpha /\beta }\right] =\int_{-\infty }^{\infty }F\left( v\right) G\left( 
\frac{t^{\alpha }}{\alpha }-v\right) dv,
\end{equation}%
which upon replacing $v=\frac{t_{1}^{\alpha }}{\alpha }$ results in 
\begin{equation}
\mathbb{F}_{\alpha /\beta }^{-1}\left[ \tilde{f}_{\alpha /\beta }\tilde{g}%
_{\alpha /\beta }\right] =\int_{\Gamma _{1}^{\alpha }}^{\infty }F\left( 
\frac{t_{1}^{\alpha }}{\alpha }\right) G\left( \frac{t^{\alpha }}{\alpha }-%
\frac{t_{1}^{\alpha }}{\alpha }\right) dt_{1}^{\alpha }.
\end{equation}%
Noting that the integrand is active for $t_{1}>t$ , this then serves as a
basis for a definition of a conformable convolution:%
\begin{equation}
f\ast g=\int_{\Gamma _{1}^{\alpha }}^{\infty }F\left( \frac{t_{1}^{\alpha }}{%
\alpha }\right) G\left( \frac{t^{\alpha }}{\alpha }-\frac{t_{1}^{\alpha }}{%
\alpha }\right) dt_{1}^{\alpha },  \label{convolutiondef}
\end{equation}%
such that 
\begin{equation}
\mathbb{F}_{\alpha /\beta }\left[ f\ast g\right] =\tilde{f}_{\alpha /\beta
}(\omega )\tilde{g}_{\alpha /\beta }(\omega ).  \label{productformula}
\end{equation}

\subsubsection{Special cases}

The transform pair defined in Eqs. (\ref{fracFLTdef}) and (\ref%
{fractFLTinvdef}) carries with it several special cases. First, $\alpha
=\beta =1,$ 
\begin{equation}
\mathbb{F}_{1/1}[f]=\int_{-\infty }^{\infty }f(t)e^{i\omega t}dt.
\end{equation}%
$\mathbb{F}_{1/1}$ is just the regular Fourier transform.

The second special case is when $\beta =\alpha .$ Then, 
\begin{equation}
\mathbb{F}_{\alpha /\beta }[f]=\int_{\Gamma ^{\alpha }}^{\infty }f(t)e^{%
\frac{i}{\alpha ^{2}}\omega ^{\alpha }t^{\alpha }}dt^{\alpha }.
\end{equation}%
This is similar to, but not exactly, the definition used by Jumarie.\cite%
{Jumarie} In that work the exponential is placed by a Mittag-Leffler
function of $\omega ^{\alpha }t^{\alpha }.$

A third case arises when $\beta =1,$ 
\begin{equation}
\mathbb{F}_{\alpha /1}[f]=\int_{\Gamma ^{\alpha }}^{\infty }f(t)e^{\frac{i}{%
\alpha }\omega t^{\alpha }}dt^{\alpha },
\end{equation}%
which is the transform used by Abdeljawad \cite{Abdel}.

The conversion to a conformable Laplace transform is trivially achieved by
replacing $i\frac{\omega ^{\beta }}{\beta }$ with $-\frac{s^{\beta }}{\beta }
$ in Eq. (\ref{fracFLTdef}). Then,%
\begin{equation}
\mathbb{L}_{\alpha /\beta }\left[ f(t)\right] =\bar{f}_{\alpha /\beta
}(s)=\int_{0}^{\infty }f(t)e^{-\frac{1}{\alpha \beta }s^{\beta }t^{\alpha
}}dt^{\alpha }.  \label{fracLTdef}
\end{equation}

Presented here are some applications of Eqs. (\ref{fracFLTdef}) and (\ref%
{fractFLTinvdef}) to particular functions. In some cases one can recover
relatively simple expressions but in other cases software such as \textsc{%
Mathematica} can compute a (complicated) solution that can be plotted.

It is of importance to consider functions that are explicit in $\frac{%
t^{\alpha }}{\alpha }.$ In this case 
\begin{equation}
\mathbb{F}_{\alpha /\beta }\left[f\left( \frac{t^{\alpha }}{\alpha }\right)
\right]=\int_{\Gamma ^{\alpha }}^{\infty }f(\frac{t^{\alpha }}{\alpha })e^{\frac{i%
}{\alpha \beta }\omega ^{\beta }t^{\alpha }}t^{\alpha -1}dt,
\end{equation}%
which upon the change of variables $u=\frac{t^{\alpha }}{\alpha }$ becomes 
\begin{equation}
\mathbb{F}_{\alpha /\beta }\left[f\left( \frac{t^{\alpha }}{\alpha }\right)
\right]=\int_{-\infty }^{\infty }f(u)e^{i\frac{\omega ^{\beta }}{\beta }u}du=%
\tilde{f}\left( \frac{\omega ^{\beta }}{\beta }\right) ,
\end{equation}%
where $\tilde{f}$ (with no subscript $\alpha /\beta $) is simply the regular
Fourier transform. Likewise,%
\begin{equation}
\mathbb{L}_{\alpha /\beta }\left[ f\left( \frac{t^{\alpha }}{\alpha }\right) %
\right] =\bar{f}\left( \frac{s^{\beta }}{\beta }\right) ,  \label{LTexplta}
\end{equation}%
where $\bar{f}$ is the regular Laplace transform. Consequently, the
conformable Laplace transform can be read directly from standard tables.\cite%
{LTtable} Table I contains a number of common transforms.

Of course, a function not explicit in $\frac{t^{\alpha }}{\alpha }$ can be
made so, $f(t)\Rightarrow F\left( \frac{t^{\alpha }}{\alpha }\right) .$ In
principle this should be of practical utility in evaluating conformable
Laplace transforms; in practice, however, one often is still confronted with
integrals that are not known. One important example are the monomials, $%
f(t)=t^{n}$. Placed in $\frac{t^{\alpha }}{\alpha }$ form this becomes%
\begin{equation}
f(t)=\alpha ^{\frac{n}{\alpha }}\left( \frac{t^{\alpha }}{\alpha }\right) ^{%
\frac{n}{\alpha }}.
\end{equation}%
So,%
\begin{eqnarray}
\mathbb{L}_{\alpha /\beta }\left[ \alpha ^{\frac{n}{\alpha }}\left( \frac{%
t^{\alpha }}{\alpha }\right) ^{\frac{n}{\alpha }}\right] &=&\mathfrak{L}%
\left[ \alpha ^{\frac{n}{\alpha }}u^{\frac{n}{\alpha }}\right] \\
&=&\alpha ^{\frac{n}{\alpha }}\beta ^{(\frac{n}{\alpha }+1)}\frac{\Gamma (%
\frac{n}{\alpha }+1)}{\left( s^{\beta }\right) ^{(\frac{n}{\alpha }+1)}} ,
\notag
\end{eqnarray}%
where $\mathfrak{L}$ is the regular Laplace transform and $\Gamma $ is the
gamma function. Table I shows this and several other examples.

\bigskip
\begin{quote}
\noindent\textbf{Table I}: Several Laplace ($\bar{f}_{\alpha /\beta }(s))$
and Fourier ($\tilde{f}_{\alpha /\beta }(\omega )$) transforms. Here 
$n=1,$ $2,$ $\cdots ,$ $p>1,$ $\Gamma $ is the gamma function, and Erfc is
the complimentary error function. For the Laplace transforms, it is assumed
that there is an appropriate Bromwich contour for the inverse operation.
The Fourier transforms require functions that $f(t)$ vanishes
sufficiently rapidly in the limit as $t\rightarrow \infty .$
\end{quote}
\bigskip

\begin{tabular}{ccc}
\hline
$f(t)$ & $\bar{f}_{\alpha /\beta }(s)$ & $\tilde{f}_{\alpha /\beta }(\omega
) $ \\ \hline\hline
$1$ & $\frac{\beta }{s^{\beta }}$ & $2\pi \delta (s^{\beta })$ \\ 
$t$ & $\alpha ^{\frac{1}{\alpha }}\beta ^{(\frac{1}{\alpha }+1)}\frac{\Gamma
(\frac{1}{\alpha }+1)}{\left( s^{\beta }\right) ^{\frac{1}{\alpha }+1}}$ & --
\\ 
$t^{n}$ & $\alpha ^{\frac{n}{\alpha }}\beta ^{(\frac{n}{\alpha }+1)}\frac{%
\Gamma (\frac{n}{\alpha }+1)}{\left( s^{\beta }\right) ^{\frac{n}{\alpha }+1}%
}$ & -- \\ 
$\frac{t^{\alpha }}{\alpha }$ & $\left( \frac{\beta }{s^{\beta }}\right)
^{2} $ & -- \\ 
$\left( \frac{t^{\alpha }}{\alpha }\right) ^{p}$ & $\left( \frac{\beta }{%
s^{\beta }}\right) ^{p+1}\Gamma (p+1)$ & -- \\ 
$e^{-k\frac{t^{\alpha }}{\alpha }}$ & $\frac{\beta }{\beta k+s^{\beta }}$ & $%
\frac{\beta }{\beta k -i\omega ^{\beta }}$ \\ 
$\left( \frac{t^{\alpha }}{\alpha }\right) ^{p}e^{-k\frac{t^{\alpha }}{%
\alpha }}$ & $\left( \frac{\beta }{\beta k+s^{\beta }}\right) ^{p+1}\Gamma
(p+1)$ & $\left( \frac{\beta }{\beta k -i\omega ^{\beta }}\right) ^{p+1}\Gamma
(p+1)$ \\ 
$\cos q\frac{t^{\alpha }}{\alpha }$ & $\frac{\beta s^{\beta }}{s^{\beta
}+\beta ^{2}q^{2}}$ & -- \\ 
$\sin q\frac{t^{\alpha }}{\alpha }$ & $\frac{\beta ^{2}q}{s^{2\beta }+\beta
^{2}q^{2}}$ & -- \\ 
$e^{-k\frac{t^{\alpha }}{\alpha }}\cos q\frac{t^{\alpha }}{\alpha }$ & $%
\frac{\left( s^{\beta }+\beta k\right) }{\left( s^{\beta }+\beta k\right)
^{2}+\beta ^{2}q^{2}}$ & $\frac{\left( \beta k-i\omega ^{\beta }\right) }{%
\left( \beta k-i\omega ^{\beta }\right) ^{2}+\beta ^{2}q^{2}}$ \\ 
$e^{-\sigma ^{2}\left( \frac{t^{\alpha }}{\alpha }\right) ^{2}}$ & $\frac{%
\sqrt{\pi }}{2\sigma }e^{\frac{s^{2\beta }}{4\sigma ^{2}\beta ^{2}}}$Erfc$%
\left[ \frac{s^{\beta }}{2\sigma \beta }\right] $ & $\frac{1 }{\sqrt{2} \sigma} e^{-\frac{1}{4 \sigma^2}\left(\frac{\omega^\beta}{\beta}\right)^2}$ \\ \hline
\end{tabular}%

\subsubsection{Physical interpretation of the transform spaces}

The Fourier transform pair of Eqs. (\ref{fracFLTdef}) and (\ref%
{fractFLTinvdef}) can generalize the concept of complimentary transform
spaces. One sees $\frac{t^{\alpha }}{\alpha }$-space is transformed to $%
\frac{\omega ^{\beta }}{\beta }$-space and vice versa in the same way the
regular Fourier transform connects $t$-space and $\omega $-space. It is
illustrative to use the applications of transforms and their spaces in
physics to help glean some insight into conformable transforms. In physical
systems time space is related to frequency space via the Fourier transform.
On the one hand, one can consider $t$ to carry units (like seconds) and $%
\omega $ to carry the inverse units. If this is the case then $\beta $ must
equal $\alpha $ such that the argument of the kernel in Eq. (\ref{fracFLTdef}%
) is unitless. Thus the $\frac{t^{\alpha }}{\alpha }$-space/$\frac{\omega
^{\beta }}{\beta }$-space connection is restricted to $\frac{t^{\alpha }}{%
\alpha }$ and $\frac{\omega ^{\alpha }}{\alpha }.$ Attempting to develop
some (albeit artificial) intuition one sees a \textquotedblleft conformable
second\textquotedblright\ in effect acting to dilate time as time goes on.
That is, a function is getting stretched out for larger values of $t.$ This
is consistent with recent work on a conformable quantum particle-in-a-box%
 \cite{Doug,Bildstein} (where instead of time, space is the independent
variable) and a classical harmonic oscillator \cite{Achar}.

One the other hand, one can start with the necessity of $t^{\alpha }\omega
^{\beta }$ to be unitless but $t$ not necessarily having units of time. Now
the $\omega ^{\beta }$ is carrying the inverse units of $t^{\alpha }$. Or,
the units on $t$ are related to that of $\omega $ as 
\begin{equation}
t=\omega ^{-\frac{\beta }{\alpha }}.
\end{equation}%
This opens up a wider relationship because $\beta $ need not equal $\alpha $
thereby connecting $t$ to a range of transform spaces. To see this, let $t$
carry conformable \textquotedblleft units\textquotedblright\ of u$^{a}$, so
that $t^{\alpha }$ has units of u$^{a\alpha }.$ Likewise let $\omega $ carry
units of u$^{-b}$, so that $\omega ^{\beta }$ has units of u$^{-b\beta }.$
The requirement that $t^{\alpha }\omega ^{\beta }$ be unitless means $\beta =%
\frac{a}{b}\alpha .$ Thus a scaling relationship exits between $\alpha $ and 
$\beta $: $\beta =\lambda \alpha .$ When $\lambda =1$, the case discussed
above ($\beta =\alpha $) is recovered and $t$ and $\omega $ have inverse
units.

As an illustrative example consider the partner functions%
\begin{equation}
f\left( t\right) =e^{-\frac{t^{\alpha }}{\alpha }}\overset{\mathbb{F}%
_{\alpha /\beta }}{\underset{\mathbb{F}_{\alpha /\beta }^{-1}}{%
\Longleftrightarrow }}f_{\alpha /\beta }(\omega )=\frac{\beta }{\beta
-i\omega ^{\beta }}.
\end{equation}%
Without a connection between $\alpha $ and $\beta $ this transform
relationship is of little utility. Consider, though, the case where $\beta
=\alpha ,$ which is plotted in Figs. 9 and 10 for $\alpha =1/4,$ $1/2,$ $3/4,
$ and 1. When $\alpha =1$, one sees the familiar Lorentzian curve for
 $\text{Re}\left[ f_{\alpha /\alpha }(\omega )\right] $ and dispersion curve for
 $\text{Im}\left[ f_{\alpha /\alpha }(\omega )\right] .$ As $\alpha $ is
decreased both curves sharpen up at low values of $\omega ,$ with very
significant compression of the curves occurring for values of $\alpha <1/2.$
Conversely, the tails of both curves for large values of $\omega $ fall away
slower for decreasing values of $\alpha .$

\begin{figure}
\includegraphics[trim=75  270 75 195,clip, width=\textwidth]{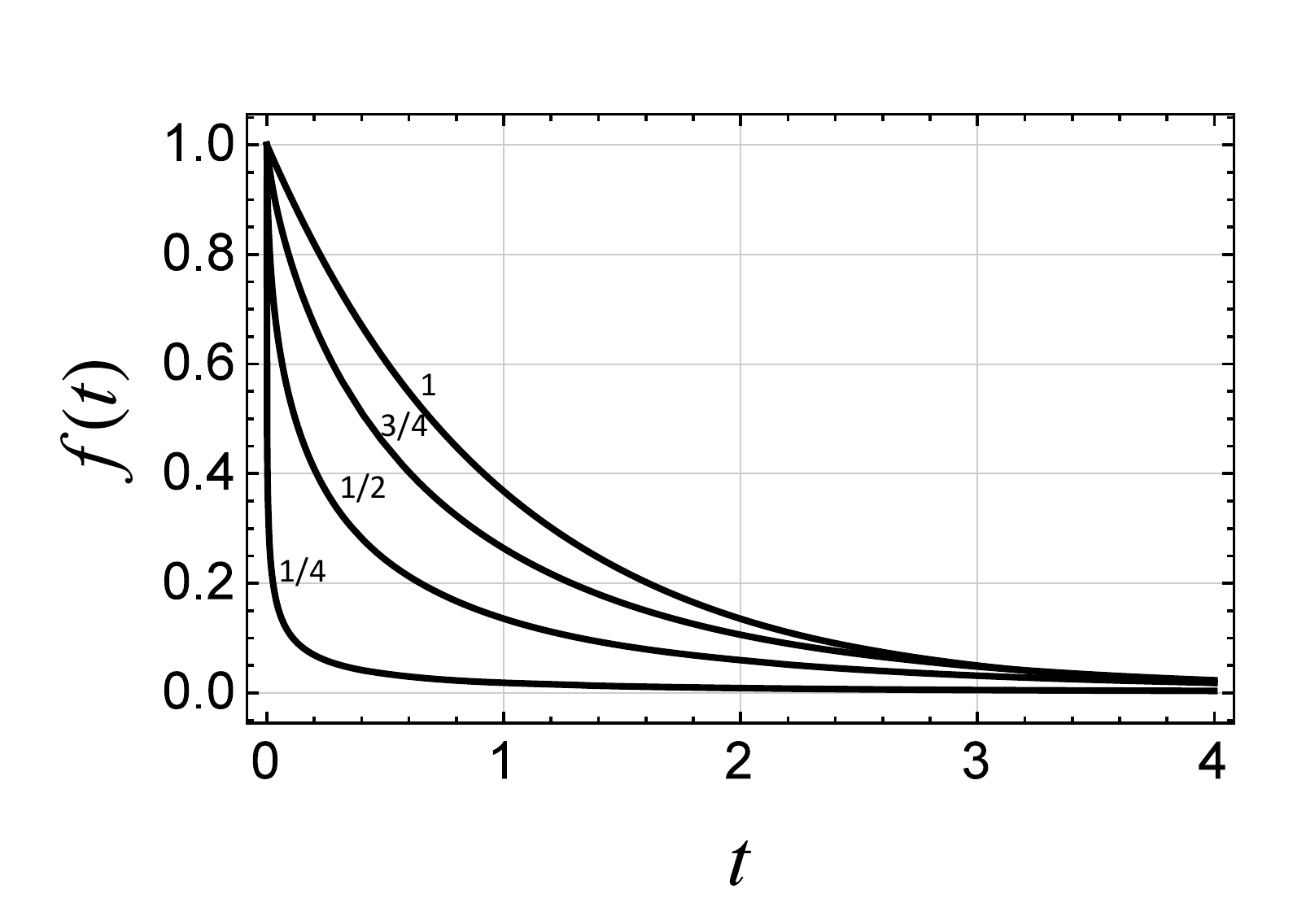}
\caption{The fractional exponential function, $f=e^{-\frac{%
t^{\alpha }}{\alpha }}$ for values of $\alpha =1/4,$ 1/2, 3/4, and 1. The
value of $\alpha =1$ yields an expoentially decreasing function. As the
value of $\alpha $ is decreased, the curve becomes much sharper. The
Fourier-Laplace transform of this function is given in Fig. 2.}
\end{figure}

\begin{figure}
\includegraphics[trim=75  225 75 50,clip, width=\textwidth]{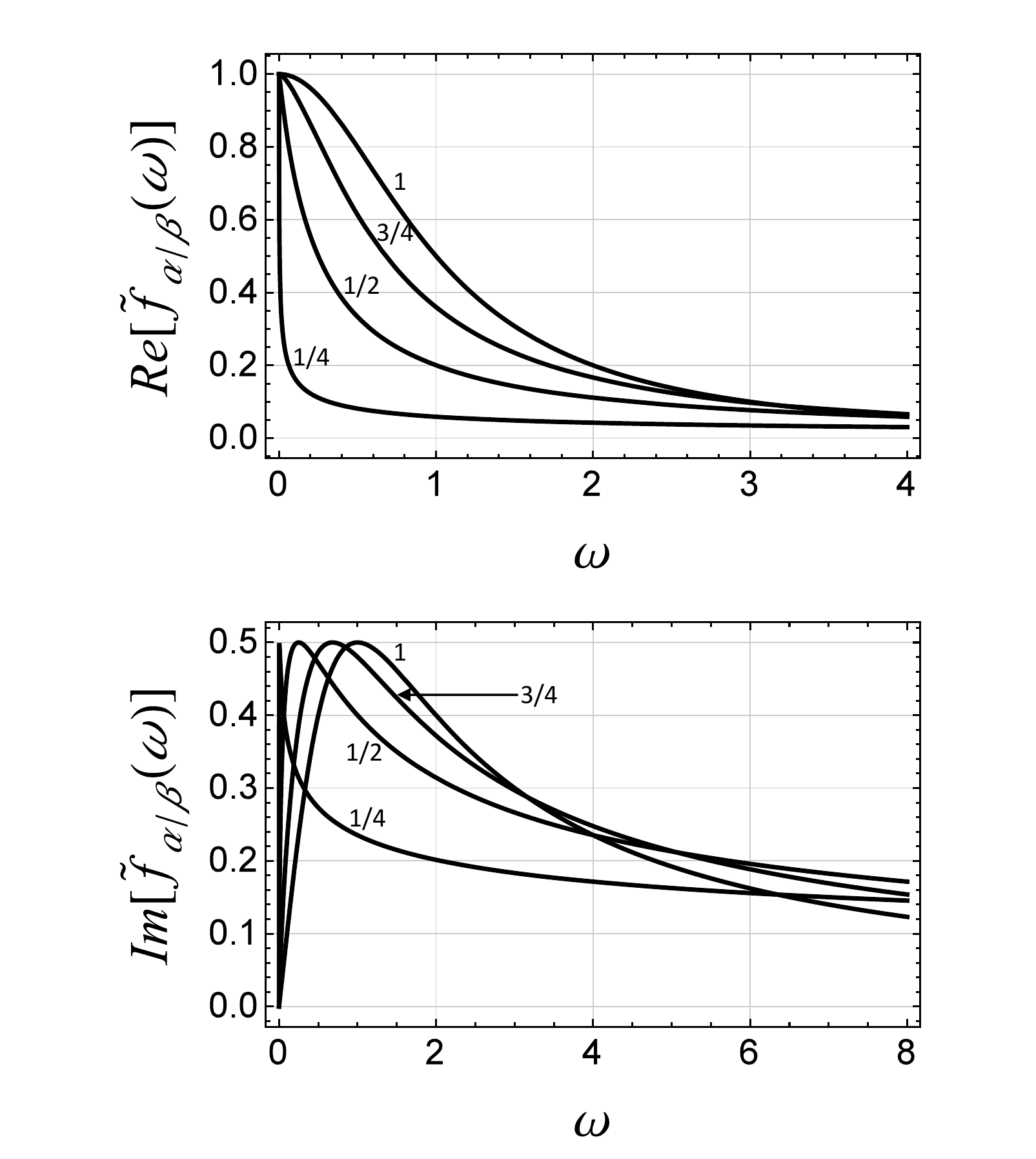}
\caption{The Fourier-Laplace transform $f_{\alpha /\beta }(\omega
)=\beta /\left( \beta -i\omega ^{\beta }\right) \ $of the fractional
exponential function, $f=e^{-\frac{t^{\alpha }}{\alpha }}$ for values of $%
\beta =1/4,$ 1/2, 3/4, and 1. The real (imaginary) part of $f_{\alpha /\beta
}(\omega )$ is plotted on the top (bottom) graph. The real part exhibits a
Lorentizan lineshape for $\beta =1$ and sharpens as $\beta $ decreases. Not
shown within this plot range, is the fact that the wings of these functions
remain elevated longer for decreasing values of $\beta .$ The imaginary part
exhibits dispersion lineshape for $\beta =1$ and again sharpens for
decreasing $\beta $ while maintaining higher values for large $\omega$.}
\end{figure}

Now consider the case where $\lambda $ varies. Here, 
\begin{equation}
f_{\alpha /\lambda \alpha }(\omega )=\frac{\lambda \alpha }{\lambda \alpha
-i\omega ^{\lambda \alpha }}.
\end{equation}%
Figure 11 shows the case where $\alpha =1$ and $\lambda =1/4,$ $1/2,$ 1, 2,
4. Here one sees a flattening of the Lorentzian curve for $\text{Re}\left[
f_{\alpha /\lambda \alpha }(\omega )\right] $ and shifting and narrowing of
the dispersion curve for $\text{Im}\left[ f_{\alpha /\lambda \alpha }(\omega
)\right] $ for values of $\lambda >1.$ For values of $\lambda <1$, the same
behavior as in Fig. 10 is seen.

\begin{figure}
\includegraphics[trim=75  225 75 50,clip, width=\textwidth]{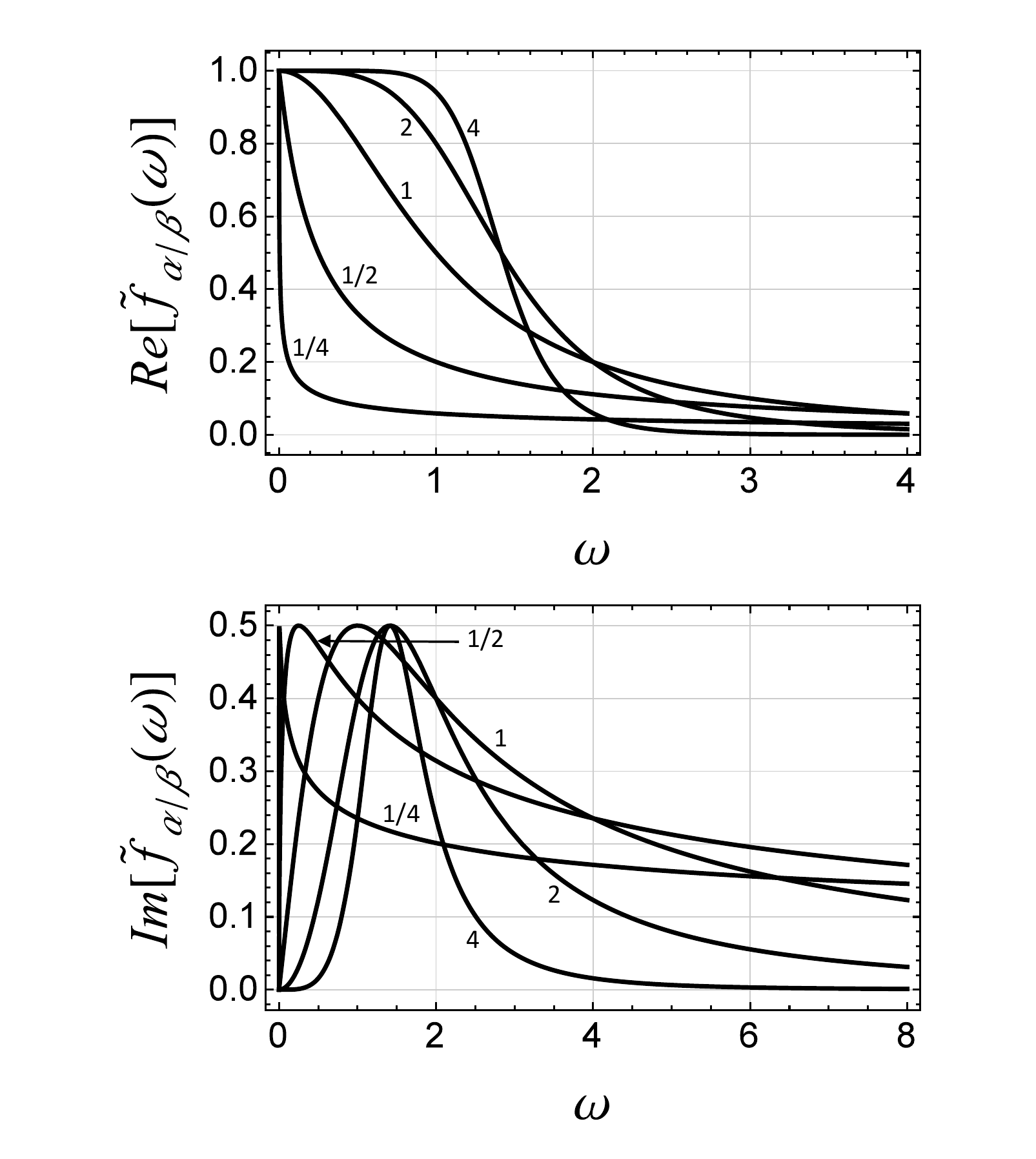}
\caption{The Fourier-Laplace transform $f_{\alpha /\beta }(\omega
)=\beta /\left( \beta -i\omega ^{\beta }\right) \ $of the fractional
exponential function when $\alpha =1$ and $\beta =\lambda \alpha $ for
values of $\lambda =1/4,$ 1/2, 1, 2, and 4. The real (imaginary) part of $%
f_{\alpha /\beta }(\omega )$ is plotted on the top (bottom) graph.
Consistent with Fig 2. The function sharpens for small values of $\omega $
but maintains higher values in the wings when $\lambda <1.$ The converse is
true for values of $\lambda >1.$}
\end{figure}

\section{Applications to quantum mechanics}

As concrete fodder for physical applications and, more importantly,
interpretation of the conformable derivative we use some examples from quantum
mechanics. We consider the conformable particle in a box and use it to
investigate conformable perturbation theory, and conformable supersymmetry.

\subsection{Conformable quantum particle in a box}

The conformable quantum particle in a box has served as a good model system
for gaining an understanding of conformable quantum mechanics.\cite%
{Bildstein,Laskin01,Laskin02,Laskin03,Laskin04,Jeng,Luchko,Guerrero,Xiao,Guo}
It has been studied using the nonlocal formulations of the conformable
derivative. This has led to controversy \cite{Jeng,Luchko,Bayin} and the
suggestion that the results for the solution to these formulations of
particle in a box cannot be valid \cite{Jeng,Luchko}. It also is difficult to
solve the problem in correct (observing nonlocality) form, although the
nonlocality itself may offer some richness to the conformable Schr\"{o}%
dinger equation \cite{Luchko}. The current work does not directly provide
input into this on-going discussion. It does however offer an alternative
formulation of the conformable quantum particle in a box that has many
appealing features. It is based upon a local formulation of the conformable
derivative (Eq. (\ref{KKder})) and it develops via Eq. (\ref{selfadjAop})
from a self-adjoint differential equation that, although complicated, is a
normal differential equation that can be solved. As such, the solutions form
an orthonormal set and the eigenvalues are real. Some issues remain with
this formulation of the conformable quantum particle in a box. Most notably
the point at $x=0$ is not a regular point.

This formulation of a conformable quantum particle in a box has been
suggested in an earlier work \cite{Doug}. We explore a few more features of
this model here; most importantly the results from perturbation theory.
Further, the concept of a \textquotedblleft phantom potential
energy\textquotedblright\ is discussed in an effort to provide some physical
insight into the model.

\subsubsection{Perturbation theory}

One can use the $\mathbb{J}_{n}^{(\alpha )}$ functions as a basis for time
independent perturbation of the particle in a box potential in the standard
way \cite{QM}. The unperturbed system is taken to be the conformable particle
in a box, with units chosen so mass and Planck's constant can be suppressed
for convenience. Then the unperturbed wavefunction is $\psi _{n}^{(0)}=%
\mathbb{J}_{n}^{(\alpha )}.$ Likewise the unperturbed energy, $E_{n}^{(0)},$
is given by Eq. (\ref{eigenvals}). The full Hamiltonian with perturbation, $%
V_{I}$ is then%
\begin{equation}
H=A_{2\alpha }+\lambda V_{I}.
\end{equation}%
Figure 12a shows the wavefunction to first order, $\psi _{1}^{(1)},$ for the
case where $V_{I}=E_{1}^{(0)}x$ and $\lambda $ is set to $-1$ (solid curve)
and 1 (dashed curve) for the case of $\alpha =1/2$. The dotted curve shows $%
\psi _{1}^{(0)}.$ Perhaps not surprisingly, when $\lambda <0$ (the potential
energy decreases with increasing $x)$ the wavefunction shifts to the right;
it becomes less skewed and more sine-like. The opposite is true when $%
\lambda >0.$ Figures 12b and 12c shows the case where 
\begin{equation}
 V_I=\left\{ \begin{array}{cc}
E_1^{(0)} & 0\leq x\leq \frac{1}{4} \\
0 & \frac{1}{4} < x \leq 1 \\
\end{array}\right. ,  \label{steppert}
\end{equation}%
for $\alpha =1/2$ and $\alpha =1/4$ respectively. The perturbation variable $%
\lambda $ is set to $-1$ (solid curve) and 1 (dashed curve). For these
cases, the presence of the step ($\lambda >0$) on the left side of the well
pushes the wavefunction to the right and it becomes more sine-like.
Conversely the presence of hole ($\lambda <0$) increases the skewing to the
left. Comparing Figs 12b and 12c one sees the impact of the perturbation is
more pronounced for $\alpha =1/4$ than for $\alpha =1/2.$

\begin{figure}
\includegraphics[trim=75  95 75 70,clip, width=\textwidth]{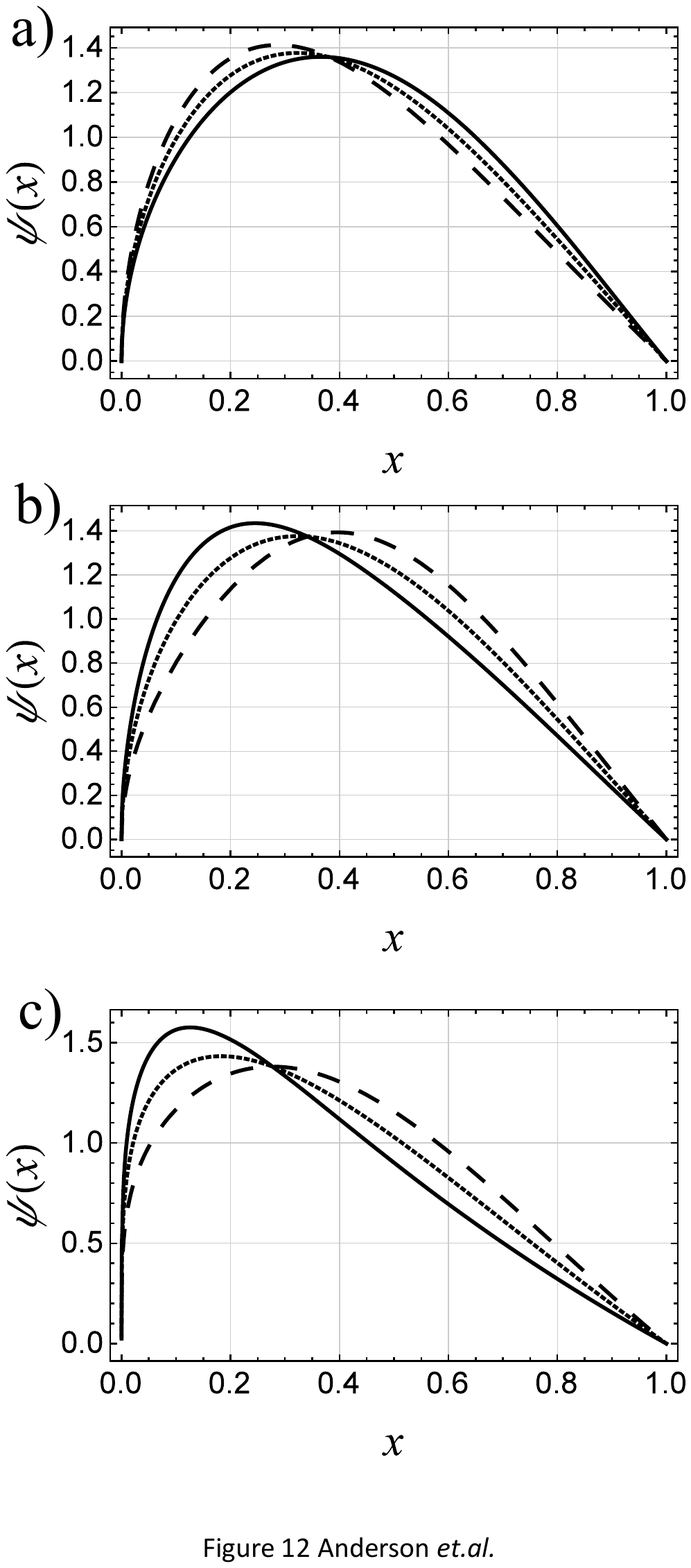}
\caption{Several representative examples of first order
wavefunctions for perturbations \textbf{(a)} $V_{I}=E_{1}^{(0)}x$ and 
\textbf{(b)} and\textbf{\ (c)} Eq. (\ref{steppert}). For (a) and (b) $\alpha
=1/2$ and for (c) $\alpha =1/4.$}
\end{figure}

\begin{figure}
\includegraphics[trim=75  225 75 150,clip, width=\textwidth]{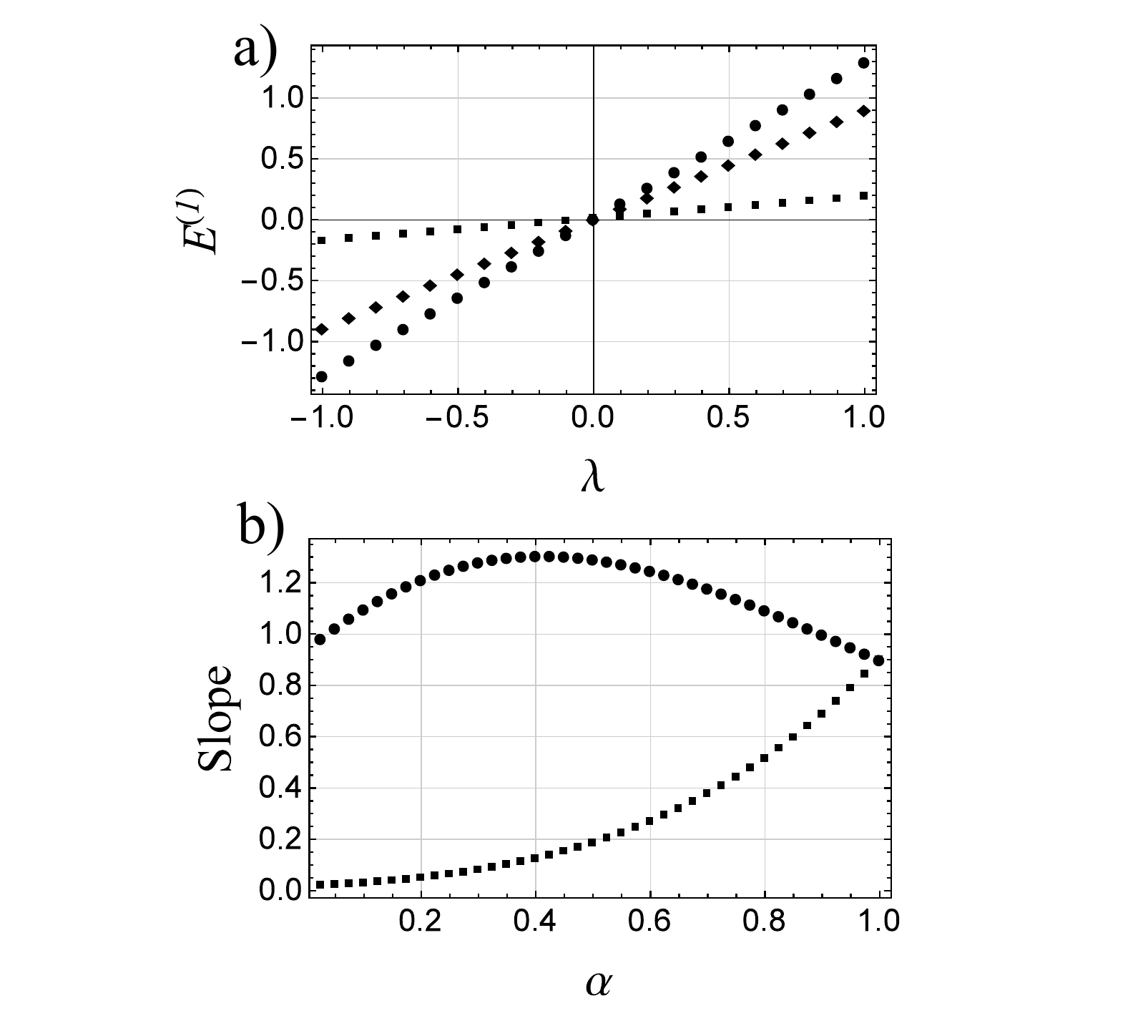}
\caption {A comparison of the effects on the first order energy
correction of a wall (or hole) on the left side (Eq. (\ref{steppert})) and
right side (\ref{steppertmir}) of the infinite box. \textbf{(a)} The ($%
\bullet $) represents the wall (or hole) on the left for $\alpha =1/2$. The (%
$\blacksquare $) represents the wall (or hole) being on the right for $%
\alpha =1/2.$ The ($\blacklozenge $) represents the comparison case for $%
\alpha =1$ which, of course, is the same for the wall (or hole) on either
side of the infinite box.\textbf{\ (b)} The slope of the line through the
points in (a) but extended to include a range of $\alpha $ values from 0 to
1. Interestingly, the wall (or hole) has maximum impact on the ground state
energy around $\alpha =2/5$ rather than for $\alpha \rightarrow 0.$}
\end{figure}

Figure 13 considers the effect of the perturbation of Eq. (\ref{steppert})
and its mirror image,%
\begin{equation}
V_I=\left\{ \begin{array}{cc}
0 & 0 \leq x \leq \frac{3}{4} \\
E_1^{(0)} & \frac{3}{4} < x\leq 1 \\
\end{array}\right.  ,  \label{steppertmir}
\end{equation}%
on the first order correction to the ground state energy. The solid diamond
symbol represents the data for the case where $\alpha =1.$ Because of the
symmetry of the wavefunctions for this case both perturbations (Eq. (\ref%
{steppert}) and Eq. (\ref{steppertmir})) have the same effect. The ground
state energy is increased for a step ($\lambda >0$) and decreased for a hole
($\lambda <0$). When one considers the case where $\alpha =1/2$ the symmetry
of the wavefunction about $x=1/2$ is broken. When the step (or hole) is on
the left (data: solid circle), the impact on the ground state energy is more
pronounced than when the step (or hole) is on the right side (data: solid
square). This makes intuitive sense based on the shape of the wavefunction,
but it also points to the idea that the lower values of $x$ carry more
weight than higher values (discussed more below). As $\alpha $ decreases
from unity, the distinction increases, but interestingly, not in a linear
fashion (Fig. 13b). The most pronounced difference between the effect of a
step (or hole) on the left versus right occurs at roughly a value of $\alpha
=2/5.$

\subsubsection{\textquotedblleft Phantom potential energy\textquotedblright\ 
}

As noted, a distinctive feature of the wavefunctions for $\alpha <1$ is the
skewing towards lower values of $x.$ This suggests a concept of a
\textquotedblleft phantom potential energy\textquotedblright\ when viewed
within $x$-space. The probability distribution ($\left\vert \psi
_{n}\right\vert ^{2}$) is not symmetrically distributed about $x=1/2$,
rather the low values of $x$ are \textquotedblleft
emphasized\textquotedblright\ more so than the higher values of $x.$ While
the potential energy is zero for $0\leq x\leq 1$, there is an apparent
presence of a \textquotedblleft phantom potential energy\textquotedblright\
pushing the probability distribution towards lower values of \thinspace $x.$
This \textquotedblleft phantom potential energy\textquotedblright\ is zero
for $\alpha =1$ and increases in effect as $\alpha \rightarrow 0.$ This
leads to the question of whether or not the effect of factorization of the
kinetic energy in a conformable Schr\"{o}dinger equation can be mapped to an
attendant potential energy term in a normal, non-conformable Schr\"{o}dinger
equation. There does not appear to be an analytically realizable solution to
a normal Schr\"{o}dinger equation that produces the $\mathbb{J}_{n}^{(\alpha
)}$ functions. However, one can consider forms of a perturbational potential
energy in the normal particle in a box equation which yield corrected
wavefunctions similar to the $\mathbb{J}_{n}^{(\alpha )}$ functions.

Using arbitrary functions to serve as a potential energy, perturbation
theory was used to approximate solutions of a non-conformable Schr\"{o}%
dinger Equation. As shown in Fig. 14, potential energies of the form $%
V_{I}=x,$ $V_{I}=x^{\alpha }$, and $V_{I}=x^{\alpha /2}$ all shift the
non-conformable wavefunction to the left, showing similar shape to the $%
\mathbb{J}_{n}^{(\alpha )}$ wavefunction given by the conformable Schr\"{o}%
dinger Equation. For Fig. 14, $\alpha =1/2.$

\begin{figure}
\includegraphics[trim=75  225 75 150,clip, width=\textwidth]{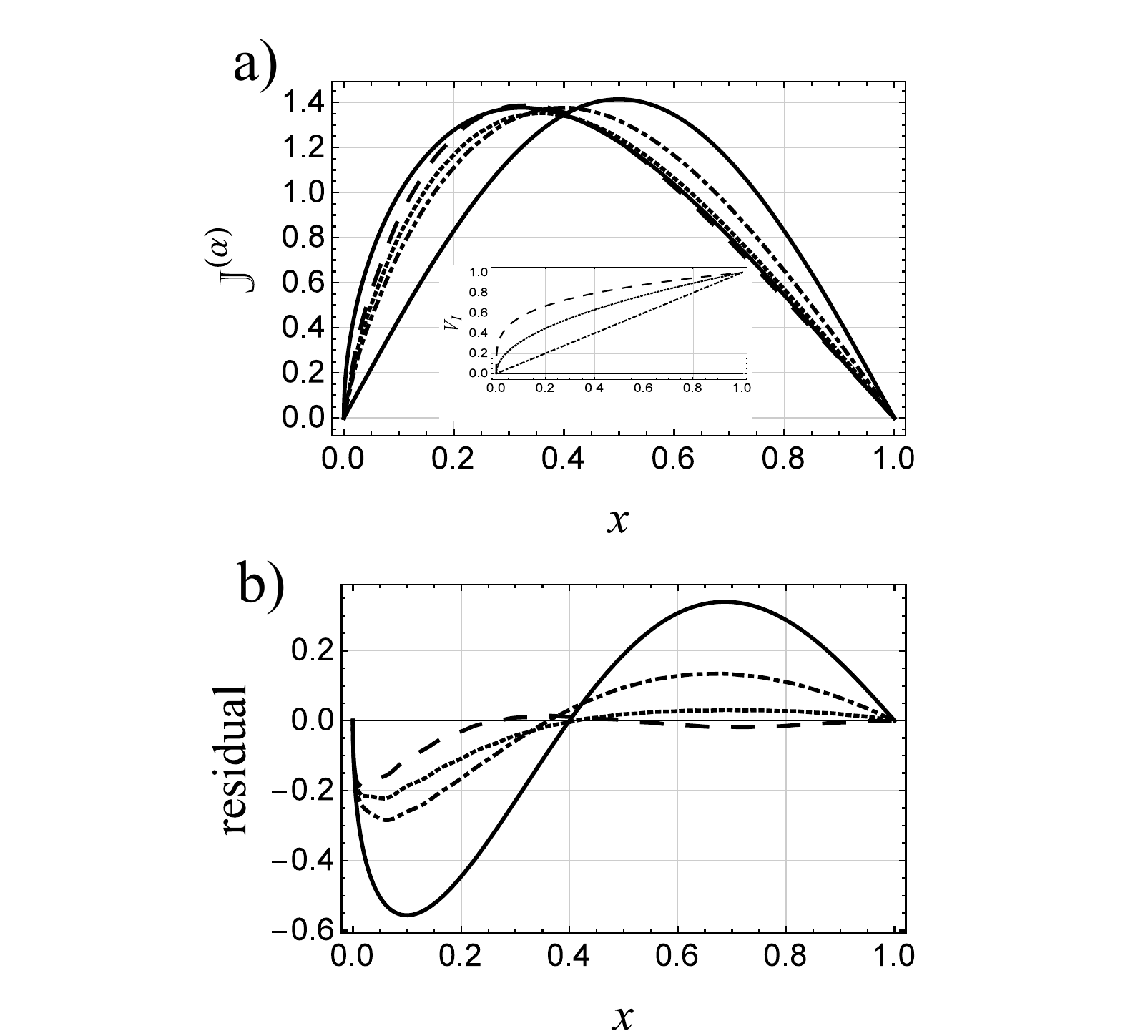}
\caption { \textbf{(a)} First order correction to the regular
particle in a box wavefunctions for $V_{I}=x$ (dash) $V_{I}=x^{\alpha }$
(dot), and $V_{I}=x^{\alpha /2}$ (dash-dot). The two solid curves are
unperturbed wavefunction, $\sqrt{2}\sin \pi x,$ and the target function, $%
\mathbb{J}_{1}^{(1/2)}(x)$. The inset shows $V_{I}.$ \textbf{(b)} The
residual curve, $\psi ^{(1)}-\mathbb{J}_{1}^{(1/2)}(x),$ for each of the
curves shown in (a).}
\end{figure}

These first-order corrected wavefunctions provide a link between the
properties of the conformable and non-conformable Schr\"{o}dinger equations.
In conformable form, the Schr\"{o}dinger equation creates a first-derivative
term, something absent from the non-conformable equation. The goal here was
to use a simple function as a perturbation to the normal particle in a box
system. Motivation for the choice of $x^{\alpha }$ resulted from the
presence of $x^{-\alpha }$ on the conformable Schr\"{o}dinger equation's
first derivative term. The $x$ and $x^{\alpha /2}$ choices served as natural
comparisons. For the particular case depicted in Fig. 14, $V_{I}=x^{\alpha
/2}$ gave the best fit to the target function, $\mathbb{J}_{n}^{1/2}$ as
seen in the residual plot, Fig 14b. For $\alpha =3/4$, $V_{I}=x$ produces
the best match (not shown).

\subsection{Shape invariance supersymmetry in conformable Sturm-Liouville
systems}

The particle in a box model is often used as a pedagogical example in
supersymmetric procedures such as SUSY \cite{SUSY1,SUSY2}. It is interesting
to do that here for the conformable particle in a box model.

\subsubsection{Symmetric differential operators}

Consider the class of generic conformable, symmetric differential operators
(note the different font is used to indicate that these operators are not
necessarily conformable derivative operators),%
\begin{equation}
\hat{p}_{S}=-i\frac{\mathcal{D}^{\alpha }+\mathcal{D}^{\beta }}{2}.
\label{pS}
\end{equation}%
Here $\mathcal{D}^{\gamma }$ where $0\leq \gamma \leq 1$ is a conformable
differential operator that satisfies the product (and quotient) rule. Note,
in general $\left[ \mathcal{D}^{\alpha },\mathcal{D}^{\beta }\right] \neq 0$
and $\mathcal{D}^{\alpha }\mathcal{D}^{\beta }\neq \mathcal{D}^{\alpha
+\beta }$. One sees that when $\alpha =\beta =1$ one recovers the regular
differential operator $\hat{p}=-i\mathcal{D}$ (when $D=\frac{d}{dx}$). We
now consider the class of conformable Sturm-Liouville systems for the form%
\begin{eqnarray}
\hat{p}_{S}^{2}\psi (x)+V(x)\psi (x) &=&\Lambda \psi (x) \\
\hat{H}\psi &=&\Lambda \psi,  \notag
\end{eqnarray}%
where $\hat{H}=\hat{p}_{S}^{2}\psi +V(x).$

We now consider the case where the set of eigenvalues has the form $%
\left\{ \Lambda _{0}>0,\Lambda _{n}\geq \Lambda _{n+1}\right\} ,$ $%
n=0,1,2,\ldots $, with corresponding eigenfunctions $\left\{ \psi
_{n}(x)\right\} .$ Following the normal shape invariance/SUSY\ procedure %
\cite{SUSY1,SUSY2,SUSY3,SUSY4,SUSY5}, we define $\hat{H}_{1}=\hat{H}-\Lambda
_{0}$ (so $V_{1}=V-\Lambda _{0}$) and consider $\hat{H}_{1}\varphi =\Lambda
^{(1)}\varphi $ for which $\hat{H}_{1}\varphi _{0}=0$ because $\Lambda
_{0}^{(1)}=0.$

Use of Eq. (\ref{pS}) and expansion gives 
\begin{equation}
-\frac{1}{4}\left( \mathcal{D}^{\alpha }\mathcal{D}^{\alpha }+\mathcal{D}%
^{\beta }\mathcal{D}^{\alpha }+\mathcal{D}^{\alpha }\mathcal{D}^{\beta }+%
\mathcal{D}^{\beta }\mathcal{D}^{\beta }\right) \varphi _{0}+V_{1}\varphi
_{0}=0.  \label{Schro1}
\end{equation}%
So, 
\begin{equation}
V_{1}=\frac{\left( \mathcal{D}^{\alpha }\mathcal{D}^{\alpha }+\mathcal{D}%
^{\beta }\mathcal{D}^{\alpha }+\mathcal{D}^{\alpha }\mathcal{D}^{\beta }+%
\mathcal{D}^{\beta }\mathcal{D}^{\beta }\right) \varphi _{0}}{4\varphi _{0}}.
\label{V1one}
\end{equation}%
We now define the following operators ($W=W(x)$) 
\begin{eqnarray}
A &=&\frac{1}{2}\left( \mathcal{D}^{\alpha }+W\right) \\
\bar{A} &=&\frac{1}{2}\left( -\mathcal{D}^{\alpha }+W\right) \\
B &=&\frac{1}{2}\left( \mathcal{D}^{\beta }+W\right) \\
\bar{B} &=&\frac{1}{2}\left( -\mathcal{D}^{\beta }+W\right) ,
\end{eqnarray}%
and consider $\left( \bar{A}+\bar{B}\right) \left( A+B\right) =\bar{A}A+\bar{%
B}A+\bar{A}B+\bar{B}B.$ The ordered products can be shown to be 
\begin{eqnarray}
\bar{A}A &=&\frac{1}{4}\left( -\mathcal{D}^{\alpha }\mathcal{D}^{\alpha }-%
\mathcal{D}^{\alpha }W+W^{2}\right) \\
\bar{B}A &=&\frac{1}{4}\left( -\mathcal{D}^{\beta }\mathcal{D}^{\alpha }+W%
\mathcal{D}^{\alpha }-W\mathcal{D}^{\beta }-\mathcal{D}^{\beta
}W+W^{2}\right) \\
\bar{A}B &=&\frac{1}{4}\left( -\mathcal{D}^{\alpha }\mathcal{D}^{\beta }+W%
\mathcal{D}^{\beta }-W\mathcal{D}^{\alpha }-\mathcal{D}^{\alpha
}W+W^{2}\right) \\
\bar{B}B &=&\frac{1}{4}\left( -\mathcal{D}^{\beta }\mathcal{D}^{\beta }-%
\mathcal{D}^{\beta }W+W^{2}\right) ,
\end{eqnarray}%
where the product rule was employed and $\mathcal{D}^{\gamma }Wf(x)$ means $%
\left( \mathcal{D}^{\gamma }W\right) f(x)$ as opposed to $\mathcal{D}%
^{\gamma }\left[ Wf(x)\right] $.

We can express $\hat{H}_{1}$ in terms of these ordered products 
\begin{eqnarray}
\hat{H}_{1} &=&\bar{A}A+\bar{B}A+\bar{A}B+\bar{B}B  \label{ABHAM} \\
&=&\frac{1}{4}\left( -\Delta +4W^{2}-2(\mathcal{D}^{\alpha }W+\mathcal{D}%
^{\beta }W)\right),  \notag
\end{eqnarray}%
where $\Delta \equiv \mathcal{D}^{\alpha }\mathcal{D}^{\alpha }+\mathcal{D}%
^{\beta }\mathcal{D}^{\alpha }+\mathcal{D}^{\alpha }\mathcal{D}^{\beta }+%
\mathcal{D}^{\beta }\mathcal{D}^{\beta }.$ Comparison of Eq. (\ref{ABHAM})
with Eq. (\ref{Schro1}) reveals 
\begin{equation}
V_{1}=W^{2}-\frac{1}{2}(\mathcal{D}^{\alpha }W+\mathcal{D}^{\beta }W).
\label{V1two}
\end{equation}%
Using Eq. (\ref{V1one}) gives a first order conformable first order
differential equations for $W.$ Guided by regular SUSY,\cite%
{SUSY1,SUSY2,SUSY3,SUSY4,SUSY5} we make the ansatz that 
\begin{equation}
W=-\frac{(\mathcal{D}^{\alpha }\varphi _{0}+\mathcal{D}^{\beta }\varphi _{0})%
}{2\varphi _{0}}.  \label{anzatz}
\end{equation}%
To confirm the ansatz we see%
\begin{equation}
W^{2}=\frac{1}{4\varphi _{0}^{2}}\left( \mathcal{D}^{\beta }\varphi _{0}%
\mathcal{D}^{\beta }\varphi _{0}+\mathcal{D}^{\beta }\varphi _{0}\mathcal{D}%
^{\alpha }\varphi _{0}+\mathcal{D}^{\alpha }\varphi _{0}\mathcal{D}^{\beta
}\varphi _{0}+\mathcal{D}^{\beta }\varphi _{0}\mathcal{D}^{\beta }\varphi
_{0}\right) ,
\end{equation}%
\begin{equation}
\mathcal{D}^{\alpha }\varphi _{0}=\frac{\mathcal{D}^{\alpha }\mathcal{D}%
^{\alpha }\varphi _{0}+\mathcal{D}^{\alpha }\mathcal{D}^{\beta }\varphi _{0}%
}{2\varphi _{0}},
\end{equation}%
and%
\begin{equation}
\mathcal{D}^{\beta }\varphi _{0}=\frac{\mathcal{D}^{\beta }\mathcal{D}%
^{\alpha }\varphi _{0}+\mathcal{D}^{\beta }\mathcal{D}^{\beta }\varphi _{0}}{%
2\varphi _{0}}.
\end{equation}%
So the right hand side of Eq. (\ref{V1two}) becomes 
\begin{eqnarray}
W^{2}-\frac{1}{2}(\mathcal{D}^{\alpha }W+\mathcal{D}^{\beta }W) &=&\frac{%
\left( \mathcal{D}^{\alpha }\mathcal{D}^{\alpha }+\mathcal{D}^{\beta }%
\mathcal{D}^{\alpha }+\mathcal{D}^{\alpha }\mathcal{D}^{\beta }+\mathcal{D}%
^{\beta }\mathcal{D}^{\beta }\right) \varphi _{0}}{4\varphi _{0}}  \notag \\
&=&V_{1}
\end{eqnarray}%
and the ansatz is confirmed.

Finding analogy with regular SUSY we consider the reverse ordered products $%
\left( A+B\right) \left( \bar{A}+\bar{B}\right) $ to construct the partner
operator, 
\begin{equation}
H_{2}=A\bar{A}+A\bar{B}+B\bar{A}+B\bar{B}.
\end{equation}%
With some analysis similar to above one obtains 
\begin{equation}
H_{2}=\frac{1}{4}\left( -\Delta +4W^{2}+2(\mathcal{D}^{\alpha }W+\mathcal{D}%
^{\beta }W)\right) .
\end{equation}%
Thus the partner potential is 
\begin{equation}
V_{2}=W^{2}+\frac{(\mathcal{D}^{\alpha }W+\mathcal{D}^{\beta }W)}{2}.
\end{equation}%
We note that when $\alpha =\beta =1$ we recover the regular SUSY partner
potentials 
\begin{equation}
\left\{ 
\begin{array}{c}
V_{1}=W^{2}-W^{\prime } \\ 
V_{2}=W^{2}+W^{\prime }%
\end{array}%
\right\}.
\end{equation}%
As with regular SUSY \cite{SUSY1,SUSY2}, the two systems are isospectral
aside from $\Lambda _{0}^{(1)},$ such that $\Lambda _{n}^{(2)}=\Lambda
_{n+1}^{(1)}.$ The wavefunctions for $H_{2},$ $\left\{ \vartheta
_{n}\right\} $ are related to those for $H_{1}$ via%
\begin{equation}
\vartheta _{n}\propto \left( A+B\right) \varphi _{n+1}
\end{equation}%
and%
\begin{equation}
\varphi _{n+1}\propto \left( \bar{A}+\bar{B}\right) \vartheta _{n}.
\end{equation}

One can confirm the appropriate anticommutor algebra by defining 
\begin{equation}
Q_{X}=\left[ 
\begin{array}{cc}
0 & 0 \\ 
X & 0%
\end{array}%
\right] ,\quad \bar{Q}_{\bar{X}}=\left[ 
\begin{array}{cc}
0 & X \\ 
0 & 0%
\end{array}%
\right] .
\end{equation}%
So, $Q=Q_{A}+Q_{B}$ and $\bar{Q}=\bar{Q}_{\bar{A}}+\bar{Q}_{\bar{B}},$ and 
\begin{eqnarray*}
\mathcal{H} &=&\left\{ \bar{Q},Q\right\} =\left[ 
\begin{array}{cc}
\left( \bar{A}+\bar{B}\right) \left( A+B\right) & 0 \\ 
0 & \left( A+B\right) \left( \bar{A}+\bar{B}\right)%
\end{array}%
\right] \\
&=&\left[ 
\begin{array}{cc}
H_{1} & 0 \\ 
0 & H_{2}%
\end{array}%
\right]
\end{eqnarray*}%
and $\left\{ \bar{Q},\bar{Q}\right\} =\left\{ Q,Q\right\} =0.$

\subsubsection{Asymmetric differential operators}

Consider the class of conformable, symmetric differential operators,%
\begin{equation}
\hat{p}=-i\mathcal{D}^{\gamma }.  \label{pasym}
\end{equation}%
The second order equation becomes 
\begin{equation}
-\mathcal{D}^{\alpha }\mathcal{D}^{\beta }\psi +V\psi =\Lambda \psi
\end{equation}%
which upon adjusting the potential energy by $\Lambda _{0}$ becomes%
\begin{equation}
-\mathcal{D}^{\alpha }\mathcal{D}^{\beta }\varphi _{0}+V_{1}\varphi _{0}=0.
\end{equation}%
Thus 
\begin{equation}
V_{1}=\frac{\mathcal{D}^{\alpha }\mathcal{D}^{\beta }\varphi _{0}}{\varphi
_{0}}.  \label{V1aysmdoub}
\end{equation}%
Now the factorization of the resultant equation is less straightforward.
Nonetheless, one can define%
\begin{eqnarray}
A &=&\mathcal{D}^{\alpha }+W_{A} \\
\bar{A} &=&-\mathcal{D}^{\alpha }+\overline{W}_{A} \\
B &=&\mathcal{D}^{\beta }+W_{B} \\
\bar{B} &=&-\mathcal{D}^{\beta }+\overline{W}_{B}.
\end{eqnarray}%
The Hamiltonian is constructed as to maintain the $\mathcal{D}^{\alpha }%
\mathcal{D}^{\beta }$ ordering of the operators, 
\begin{eqnarray}
H_{1} &=&\bar{A}B \\
H_{1} &=&\left( -\mathcal{D}^{\alpha }+\overline{W}_{A}\right) \left( 
\mathcal{D}^{\beta }+W_{B}\right)  \notag \\
H_{1} &=&-\mathcal{D}^{\alpha }\mathcal{D}^{\beta }+\overline{W}_{A}\mathcal{%
D}^{\beta }-W_{B}\mathcal{D}^{\alpha }-\mathcal{D}^{\alpha }W_{B}+\overline{W%
}_{A}W_{B},  \notag
\end{eqnarray}%
where as before $\mathcal{D}^{\alpha }W_{B}$ means $\left( \mathcal{D}%
^{\alpha }W_{B}\right) .$ The first SUSY\ partner potential is now more
complicated%
\begin{equation}
V_{1}=\overline{W}_{A}W_{B}-\mathcal{D}^{\alpha }W_{B}-W_{B}\mathcal{D}%
^{\alpha }+\overline{W}_{A}\mathcal{D}^{\beta }.
\end{equation}%
However, one can impose the condition $-W_{B}\mathcal{D}^{\alpha }+\overline{%
W}_{A}\mathcal{D}^{\beta }=0$ which will determine the functional
relationship between $W_{B}$ and $\overline{W}_{A}.$ This then leaves an
expression that is similar to regular SUSY, 
\begin{equation}
V_{1}=\overline{W}_{A}W_{B}-\mathcal{D}^{\alpha }W_{B}.  \label{V1asym}
\end{equation}%
The second SUSY\ partner Hamiltonian is again produced by requiring the $%
\mathcal{D}^{\alpha }\mathcal{D}^{\beta }$ ordering. The means $H_{2}=A\bar{B%
}$ and thus, 
\begin{equation}
V_{2}=\overline{W}_{A}W_{B}-\mathcal{D}_{B}^{\alpha }\overline{W}_{B}-W_{A}%
\mathcal{D}^{\beta }+\overline{W}_{B}\mathcal{D}^{\alpha }.
\end{equation}%
One can again require $-W_{A}\mathcal{D}^{\beta }+\overline{W}_{B}\mathcal{D}%
^{\alpha }=0$ to determine the functional relationship between $W_{A}$ and $%
\overline{W}_{B}.$ This gives 
\begin{equation}
V_{2}=\overline{W}_{A}W_{B}-\mathcal{D}_{B}^{\alpha }\overline{W}.
\end{equation}%
The asymmetric treatment is convenient for Sturm-Liouville equations. Some
examples are given below.

One important case is when $\beta =\alpha <1.$ Here,%
\begin{equation}
H_{1}=-\mathcal{D}^{\alpha }\mathcal{D}^{\alpha }\varphi +V_{1}\varphi
=\Lambda ^{(1)}\varphi  \label{SUSYH1}
\end{equation}%
and 
\begin{equation}
H_{2}=-\mathcal{D}^{\alpha }\mathcal{D}^{\alpha }\vartheta +V_{2}\vartheta
=\Lambda ^{(2)}\vartheta .  \label{SUSYH2}
\end{equation}%
Where, like regular SUSY\ $\Lambda _{0}^{(2)}=\Lambda _{1}^{(1)},$ $\Lambda
_{n}^{(2)}=\Lambda _{n+1}^{(1)}$, the partner potentials are%
\begin{equation}
\left\{ 
\begin{array}{c}
V_{1}=W^{2}-\mathcal{D}^{\alpha }W \\ 
V_{2}=W^{2}+\mathcal{D}^{\alpha }W%
\end{array}%
\right\}  \label{alphapartners}
\end{equation}%
and one sees an identical structure compared to that of regular SUSY except
with the conformable differential operator playing the role of the regular derivative.

\subsubsection{The conformable derivative.}

Consider here the case of the conformable derivative. Now $\mathcal{D}^{\alpha
}[f]=D^{\alpha }\left[ f\right] =x^{1-\alpha }f^{\prime }.$ Consider the
case for $V=0$ for $0\leq x\leq 1$ and boundary conditions $\psi (0)=\psi
(1)=0.$ 
\begin{eqnarray}
H\psi  &=&\Lambda \psi  \\
-D^{\alpha }D^{\alpha }\psi  &=&\Lambda \psi 
\end{eqnarray}%
has been studied and has solutions \cite{Doug}%
\begin{equation}
\psi _{n}=N\sin \left( \sqrt{\Lambda _{n}}\frac{x^{\alpha }}{\alpha }\right)
,
\end{equation}%
where $N$ is a normalization constant and $\Lambda _{n}=\alpha
^{2}(n+1)^{2}\pi ^{2},$ $n=0,1,2,\ldots $ Thus $\Lambda _{n}^{(1)}=\alpha
^{2}(n+1)^{2}\pi ^{2}-\alpha ^{2}\pi ^{2}=\alpha ^{2}n(n+2)\pi ^{2}$ and 
\begin{equation}
\varphi _{n}=N\sin \left( (n+1)\pi x^{\alpha }\right) .
\end{equation}%
Following regular SUSY, $\varphi _{0}=N\sin \left( \pi x^{\alpha }\right) $.
Thus from Eq. (\ref{anzatz}) with $\beta =\alpha $, 
\begin{eqnarray}
W &=&-\frac{D^{\alpha }\varphi _{0}}{\varphi _{0}}=-\frac{x^{1-\alpha
}\varphi _{0}^{\prime }}{\varphi _{0}}  \notag \\
&=&-\frac{N\alpha \pi \cos (\pi x^{\alpha })}{N\sin \left( \pi x^{\alpha
}\right) }  \notag \\
&=&-\alpha \pi \cot (\pi x^{\alpha }).
\end{eqnarray}%
The partner potential is obtained from Eq. (\ref{alphapartners}) as 
\begin{eqnarray}
V_{2} &=&\alpha ^{2}\pi ^{2}\left( \cot ^{2}(\pi x^{\alpha })+\csc ^{2}(\pi
x^{\alpha })\right)   \notag \\
&=&\alpha ^{2}\pi ^{2}\left( 2\csc ^{2}(\pi x^{\alpha })-1\right) .
\end{eqnarray}%
\bigskip The $\vartheta _{n}$ are obtained by acting on $\varphi _{n}$ with
the operator $A+B$ \ with $\beta =\alpha $ 
\begin{eqnarray}
\vartheta _{n} &\propto &\left( A+B\right) \varphi _{n+1} \\
&\propto &\left( D^{\alpha }+W\right) \varphi _{n+1}  \notag \\
&\propto &x^{1-\alpha }\varphi _{n+1}^{\prime }-\alpha \pi \cot (\pi
x^{\alpha })\varphi _{n+1}  \notag \\
&\propto &(n+2)\alpha \pi \cos \left( (n+2)\pi x^{\alpha }\right) -\alpha
\pi \cot (\pi x^{\alpha })\sin \left( (n+2)\pi x^{\alpha }\right) .  \notag
\end{eqnarray}%
The first couple of eigenstates are (with the use of some trigonometric
identities) 
\begin{equation}
\vartheta _{0}\propto \sin ^{2}(\pi x^{\alpha })
\end{equation}%
and 
\begin{equation}
\vartheta _{1}\propto \sin (\pi x^{\alpha })\sin (2\pi x^{\alpha }).
\end{equation}%
Figure 15 shows the a plot of the eigenvalue/eigenfunction system associated
with $H_{1}$ and $H_{2}$ for $\alpha =1,$ $3/4,$ 1/2, 1/4.

\begin{figure}
\includegraphics[trim=75  225 75 100,clip, width=\textwidth]{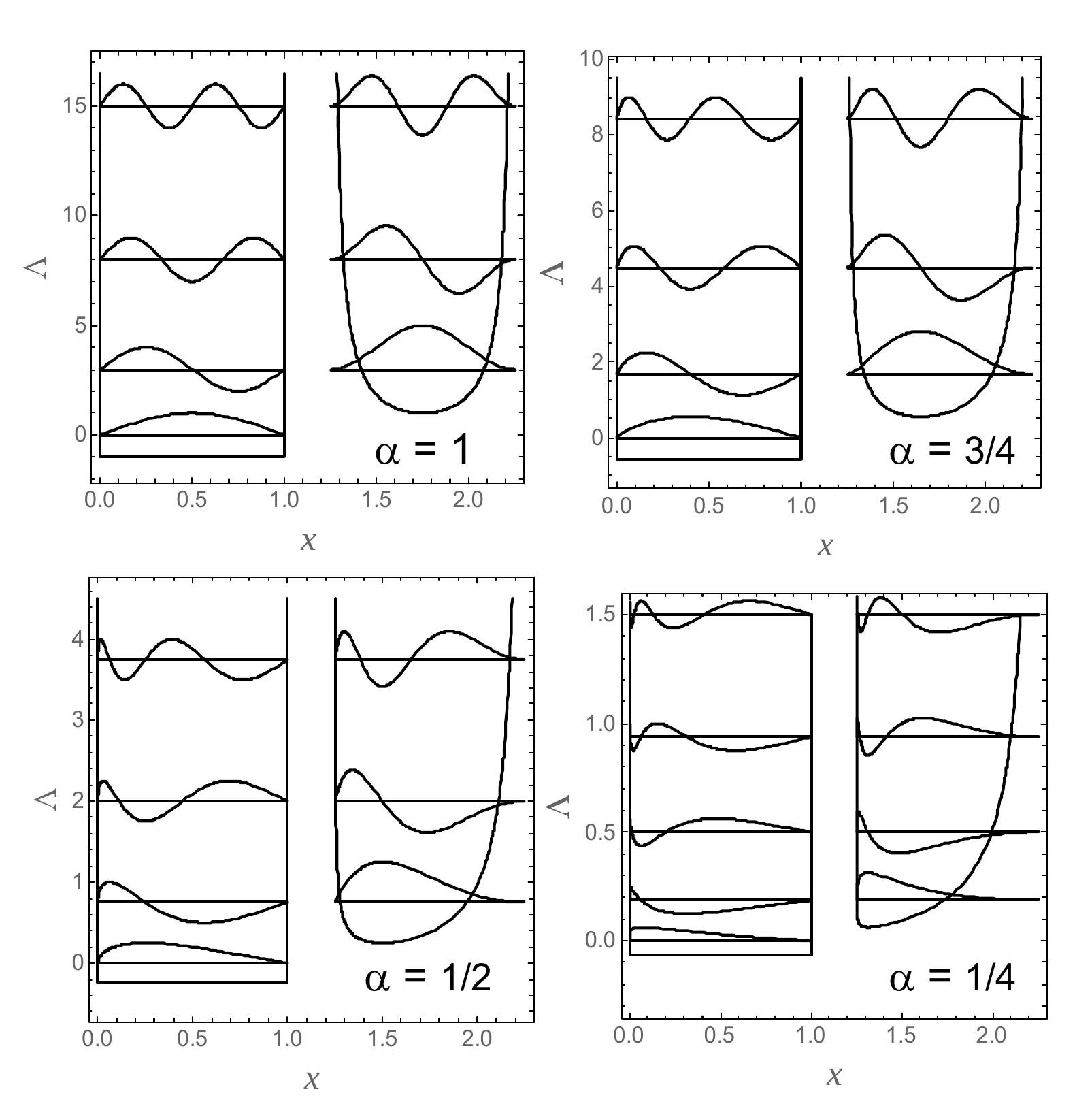}
\caption {Symmetric version of the conformable SUSY\ plot of the
particle in a box system associated with $H_{1}$ of Eq. (\ref{SUSYH1}) and $%
H_{2}$ of Eq (\ref{SUSYH2}) for $\beta =\alpha =1,$ $3/4,$ 1/2, 1/4. The
case when $\alpha =1$ recovers the regular SUSY partner potential and enegry
levels for regular SUSY \cite{SUSY1,SUSY2}}.
\end{figure}

An interesting variant is to consider an asymmetric version%
\begin{eqnarray}
\mathcal{D}^{\alpha } &=&D^{1}=\frac{d}{dx} \\
\mathcal{D}^{\beta } &=&D^{\alpha }=x^{1-\alpha }\frac{d}{dx}  \notag
\end{eqnarray}%
\begin{eqnarray}
D^{\alpha }D^{\beta } &=&\frac{d}{dx}\left[ x^{1-\alpha }\frac{d}{dx}\right]
\\
&=&x^{1-\alpha }\frac{d^{2}}{dx^{2}}+(1-\alpha )x^{-\alpha }\frac{d}{dt}. 
\notag
\end{eqnarray}%
This is a self-adjoint version of the conformable double-derivative given in Eq. (%
\ref{selfadjAop}). The solution set of 
\begin{equation}
-\frac{d}{dx}\left[ x^{1-\alpha }\frac{d}{dx}\right] \psi =\Lambda \psi
\end{equation}%
subject to $\psi (0)=\psi (1)=0,$ is (\emph{cf}., Eq. (\ref{bigJ})) 
\begin{equation}
\psi _{n}=\mathbb{J}_{n}^{(\alpha )}(x).
\end{equation}
The denominator of $\mathbb{J}_{n}^{(\alpha )}(x)$, will be important. It is
defined here as $N_{n},$ and is (\emph{cf}., Eq. (\ref{bigJ})) 
\begin{equation}
N_{n}=\frac{1}{\sqrt{\left( \eta -1\right) J_{\eta -1}\left( n_{\eta
}\right) J_{\eta +1}\left( n_{\eta }\right) }},
\end{equation}%
where $n_{\eta }$ is the $n^{th}$ zero of $J_{\eta }(x)$ for $\eta =\frac{%
\alpha }{1+\alpha }$. These wavefunctions are purely real. The energy levels
are (\emph{cf}., Eq. (\ref{eigenvals})) 
\begin{equation}
\Lambda _{n}=\frac{(1+\alpha )^{2}n_{\eta }^{2}}{4}.  \label{PIBEn}
\end{equation}

One must first solve 
\begin{eqnarray}
-W_{B}D^{\alpha }+\overline{W}_{A}D^{\beta } &=&0  \notag \\
-W_{B}\frac{d}{dx}+\overline{W}_{A}x^{1-\alpha }\frac{d}{dx} &=&0  \notag \\
-W_{B}+\overline{W}_{A}x^{1-\alpha } &=&0
\end{eqnarray}%
to obtain 
\begin{equation}
\overline{W}_{A}=x^{\alpha -1}W_{B}.  \label{ABrel1}
\end{equation}

Considering $H_{1}$ such that $\Lambda _{0}^{(1)}=0,$ leads to 
\begin{equation}
\varphi _{n}=N_{n+1}\sqrt{x^{\alpha }}J_{\eta }\left( (n+1)_{\eta
}(x^{\alpha })^{\frac{1}{2\eta }}\right) .
\end{equation}%
From Eqs. (\ref{V1asym}) and (\ref{ABrel1}), 
\begin{equation}
V_{1}=x^{\alpha -1}W_{B}^{2}-\frac{dW_{B}}{dx}
\end{equation}%
and from Eq. (\ref{V1aysmdoub})%
\begin{equation}
V_{1}=\frac{\frac{d}{dx}\left[ x^{1-\alpha }\frac{d}{dx}\right] \varphi _{0}%
}{\varphi _{0}}.
\end{equation}%
Taking the ansatz,%
\begin{equation}
W_{B}=-\frac{x^{1-\alpha }\varphi _{0}^{\prime }}{\varphi _{0}},
\end{equation}%
one may verify,%
\begin{eqnarray}
V_{1} &=&x^{\alpha -1}\left( \frac{x^{1-\alpha }\varphi _{0}^{\prime }}{%
\varphi _{0}}\right) ^{2}+\frac{-x^{1-\alpha }\left( \varphi _{0}^{\prime
}\right) ^{2}+\varphi _{0}\left( x^{1-\alpha }\varphi _{0}^{\prime \prime
}+\left( 1+\alpha \right) x^{\alpha }\varphi _{0}^{\prime }\right) }{\varphi
_{0}^{2}}  \notag \\
V_{1} &=&\frac{x^{1-\alpha }\varphi _{0}^{\prime \prime }+\left( 1+\alpha
\right) x^{\alpha }\varphi _{0}^{\prime }}{\varphi _{0}}  \notag \\
V_{1} &=&\frac{\frac{d}{dx}\left[ x^{1-\alpha }\frac{d}{dx}\right] \varphi
_{0}}{\varphi _{0}}.
\end{eqnarray}

Considering the SUSY partner potential, we again first address 
\begin{eqnarray*}
-W_{A}D^{\beta }+\overline{W}_{B}D^{\alpha } &=&0 \\
-W_{A}x^{1-\alpha }\frac{d}{dx}+\overline{W}_{B}\frac{d}{dx} &=&0 \\
-W_{A}x^{1-\alpha }+\overline{W}_{B} &=&0,
\end{eqnarray*}%
thus 
\begin{equation}
W_{A}=x_{B}^{\alpha -1}\overline{W}  \label{ABrel2}
\end{equation}%
so, 
\begin{equation}
V_{2}=x_{B}^{\alpha -1}\overline{W}_{B}^{2}+\frac{d\overline{W}_{B}}{dx}.
\end{equation}%
There is the flexibility with Eqs. (\ref{ABrel1}) and (\ref{ABrel2}) to set 
\begin{equation}
\overline{W}_{B}=W_{B}=W=-\frac{x^{1-\alpha }\varphi _{0}^{\prime }}{\varphi
_{0}}.  \label{SUSYW}
\end{equation}%
Consequently, 
\begin{eqnarray}
\vartheta _{n} &\propto &B\varphi _{n+1} \\
\vartheta _{n} &\propto &x^{\frac{1-\alpha }{2}}\frac{d\sqrt{x^{\alpha }}%
J_{\eta }\left( n_{\eta }(x^{\alpha })^{\frac{1}{2\eta }}\right) }{d}-\frac{%
x^{\frac{1-\alpha }{2}}\varphi _{0}^{\prime }}{\varphi _{0}}\sqrt{x^{\alpha }%
}J_{\eta }\left( n_{\eta }(x^{\alpha })^{\frac{1}{2\eta }}\right) .  \notag
\end{eqnarray}%
Both $W$ and $\vartheta _{n}$ are complicated combinations of Bessel
functions but can be plotted as shown in Fig. 16 which shows $W$ for $\alpha
=1,$ $3/4,$ 1/2, 1/4 and Fig 17 which is analogous to Fig 15.

\begin{figure}
\includegraphics[trim=75  350 75 150,clip, width=\textwidth]{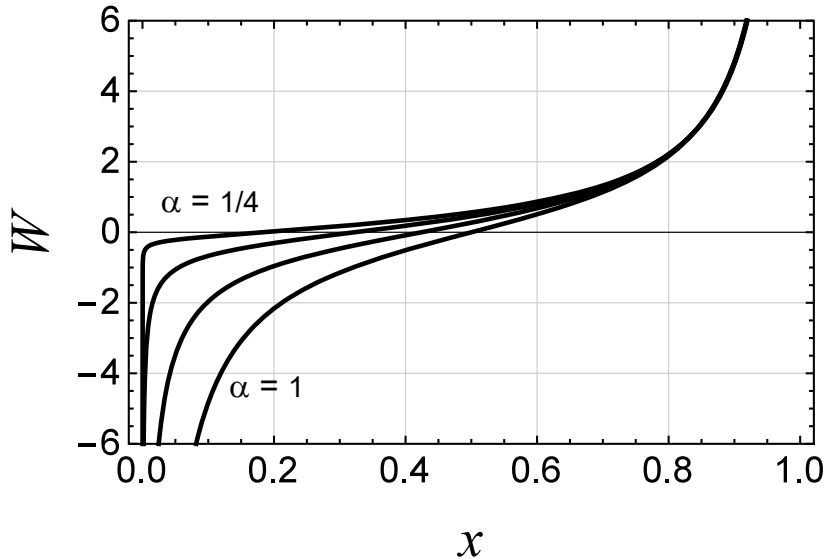}
\caption {Plot of the $W$ from Eq. (\ref{SUSYW}) for an asymmetric
version of the form $D^{\alpha }D^{\beta }=\frac{d}{dx}\left[ x^{1-\alpha }%
\frac{d}{dx}\right] $ with $\alpha =1,$ $3/4,$ 1/2, 1/4.}
\end{figure}

\begin{figure}
\includegraphics[trim=75  200 75 125,clip, width=\textwidth]{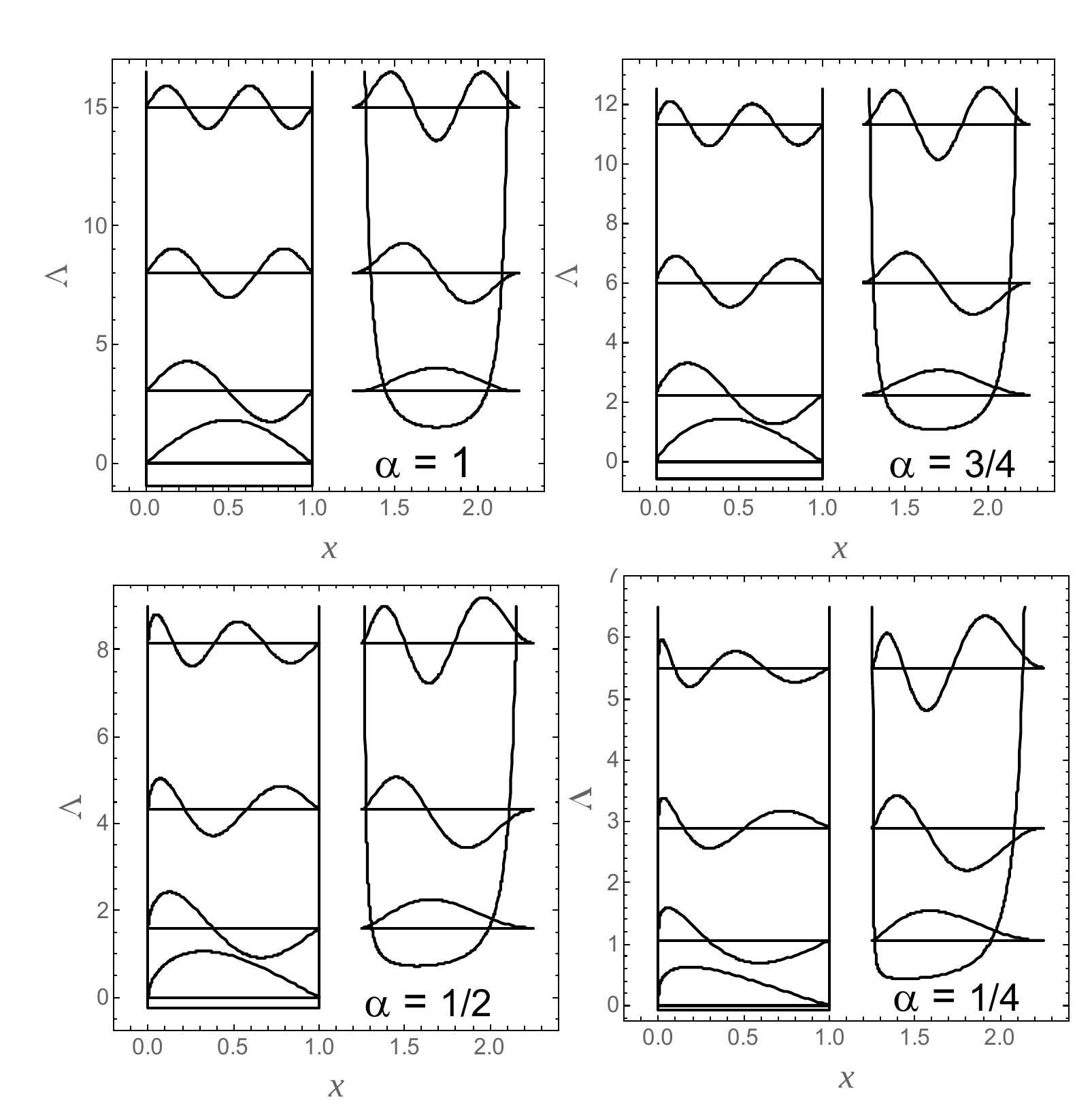}
\caption {Analog of Fig. 15 for an asymmetric version of the form $%
D^{\alpha }D^{\beta }=\frac{d}{dx}\left[ x^{1-\alpha }\frac{d}{dx}\right] $.
Plots are show for $\alpha =1,$ $3/4,$ 1/2, 1/4. Differences between the
plots show in this figure and that of Fig. 15 become more pronounced for
smaller values of $\alpha .$}
\end{figure}

\section{Conclusion}

It is hoped this work clearly shows that \emph{the conformable derivative for
differentiable functions is equivalent to a simple change of variable.} But
it is also hoped that the variety of areas shown in this work suggests that
there is value in studying the properties of the conformable derivative.

This work discussed the use of a self-adjoint operator, $\hat{A}_{2\alpha },$
which is built from the conformable derivative. The solution to the eigenvalue
problem with boundary conditions $y(0)=y(1)=0$ leads to the complete
orthonormal set of functions $\mathbb{J}_{n}^{(\alpha )}$ which are
parameterized by the $\alpha$ of the conformable derivative used
in $\hat{A}_{2\alpha }.$ Various properties of the $\mathbb{J}_{n}^{(\alpha
)}$ functions were explored including the nature of the roots of the
functions, scaling relations, and areas between zeros. The behavior of the
moments of the functions were plotted and discussed. The recasting of $%
\mathbb{J}_{n}^{(\alpha )}$ in terms of the confluent hypergeometric
limiting functions was done. The $\mathbb{J}_{n}^{(\alpha )}$ functions form
the basis for a generalization of the Fourier series and several example
functions were investigated. The relationship to the Fourier-Bessel series
was found and, although most often one would need to resort to numerical
integration, several special cases yielded analytic representation.

This work offers a fairly general consideration of the conformable Fourier
transform pair that connects $\frac{t^{\alpha }}{\alpha }$-space to $\frac{%
\omega ^{\beta }}{\beta }$-space in the same way the regular Fourier
transform connects $t$-space to $\omega $-space. This definition was shown
to be a one-to-one transform and exhibits many important properties of a
regular Fourier transform. These include the convolution theorem, formulas
for the derivative, and explicit functions of $\frac{t^{\alpha }}{\alpha }$
and/or $\frac{\omega ^{\beta }}{\beta }.$ Further, it provided a natural
framework for an expression for a conformable convolution. It is hoped that
insights into the nature of conformable derivatives and in the relationship
between $\frac{t^{\alpha }}{\alpha }$-space to $\frac{\omega ^{\beta }}{%
\beta }$-space can be gleaned from the transform pair. One can envision
potential application wherever there is a physical connection between
complementary spaces. In particular, in quantum mechanics position and
momentum are related to one another via Fourier transformation and the
physical operator representing momentum is essentially the derivative with
respect to position.

Finally this work discussed several applications in quantum mechanics.
Perturbation theory was discussed and the concept of a \textquotedblleft
phantom potential energy\textquotedblright\ was developed. As a second
application, a simple SUSY\ calculation was performed for the particle in a
box model. Perhaps the use of the conformable derivative will be valuable in forming
phenomenological models. Only quantum mechanics was discussed in this work,
but one could envision exploring other areas of physics as well.

\section*{acknowledgements}

We are thankful for support from the Concordia College Chemistry Alumni Research Fund.

\end{document}